\documentclass[prb,preprint,showpacs,preprintnumbers,amsmath,amssymb]{revtex4-1}

\usepackage{graphicx}
\usepackage{bm}% bold math
\usepackage{gensymb}
\usepackage{lineno}
\newcommand{\be}{\begin{equation}}
\newcommand{\ee}{\end{equation}}
\newcommand{\eq}[1]{Eq.~(\ref{#1})}
\newcommand{\fig}[1]{Fig.~\ref{#1}}
\def\bea{\begin{eqnarray}}
\def\eea{\end{eqnarray}}

\def\bra{\langle}
\def\ket{\rangle}
\def\vq{{\bf q}}

\def\vk{{\bf k}}

\begin{document}

\title{Fermi surface in La-based cuprate superconductors from Compton scattering imaging} 

\author{Hiroyuki Yamase$^{1,2*}$, Yoshiharu Sakurai$^{3}$, 
Masaki Fujita$^{4}$, Shuichi Wakimoto$^{5}$, and Kazuyoshi Yamada$^{6}$}
\affiliation{
{$^{1}$}International Center for Materials Nanoarchitectonics, National Institute for Materials Science (NIMS), Tsukuba 305-0047, Japan \\
{$^{2}$}Department of Condensed Matter Physics, Graduate School of Science, Hokkaido University, Sapporo 060-0810, Japan\\
{$^{3}$}Japan Synchrotron Radiation Research Institute (JASRI), SPring-8, Hyogo 679-5198, Japan \\
{$^{4}$}Institute for Materials Research, Tohoku University, Sendai 980-8577, Japan \\
{$^{5}$}Materials Sciences Research Center, Japan Atomic Energy Agency, Tokai, Naka, Ibaraki 319-1195, Japan \\
{$^{6}$} High Energy Accelerator Research Organization (KEK), Tsukuba 305-0801, Japan
}

\date{20 February, 2020}

\begin{abstract}
Compton scattering provides invaluable information on the underlying Fermi surface (FS) 
and is a powerful tool complementary to angle-resolved photoemission spectroscopy 
and quantum oscillation measurements. 
Here we perform high-resolution Compton scattering 
measurements for La$_{2-x}$Sr$_{x}$CuO$_{4}$ with $x=0.08$ ($T_c=20$~K) 
at $300$~K and 150~K, and image the momentum distribution function 
in the two-dimensional Brillouin zone. 
We find that the observed images cannot be reconciled with the conventional hole-like FS believed so far. 
Instead, our data imply that the FS is strongly deformed by the underlying nematicity in each CuO$_2$ plane, 
but the bulk FSs recover the fourfold symmetry. 
We also find an unusually strong temperature dependence of the momentum distribution function, 
which may originate from the pseudogap formation 
in the presence of the reconstructed FSs due to the underlying nematicity. 
Additional measurements for $x=0.15$ and $0.30$ at $300$~K suggest 
similar FS deformation with weaker nematicity, which nearly vanishes at $x=0.30$.

%\vspace{10mm}

%\noindent Correspondence to: yamase.hiroyuki@nims.go.jp
\end{abstract}

\maketitle
\section{introduction} 
The Fermi surface (FS) is a direct consequence of the Pauli exclusion principle 
and is very fundamental in the condensed matter physics. In particular, 
the shape of the FS reflects the electron motion inside a material and is a key to control 
the low-energy properties of metals. 

High-temperature cuprate superconductors are one of interesting metals and exhibit 
various phenomena by changing carrier doping and temperature \cite{keimer15}: 
the pseudogap phase, nematic order, charge-density-wave, 
spin-density-wave, and superconductivity. While no consensus has been 
obtained on the coherent understanding of those phenomena, it is well recognized 
that the FS plays a major role to understand the complicated physics in cuprates. 
However, the underlying FS in high-temperature cuprate 
superconductors remains elusive and controversial. 

By applying a high magnetic field and suppressing the superconductivity, 
quantum oscillation measurements revealed that the large FS in the overdoped 
region is reconstructed to small Fermi pockets in the underdoped region  
at low temperature \cite{sebastian15}. 
This reconstruction was interpreted to be driven by 
translation symmetry breaking due to a charge-density-wave instability 
in the magnetic field. 
Recent Hall number measurements \cite{badoux16} suggested that 
the charge-density-wave scenario alone cannot be a whole story.  
The FS reconstruction can be a consequence of both spiral (not collinear \cite{charlebois17})  
magnetic order and charge-density-wave in the high field \cite{eberlein16}. 
Angle-resolved photoemission spectroscopy (ARPES) can reveal the FS 
in a condition without a magnetic field \cite{damascelli03}. 
A clear signature of the reconstruction 
of the FS is not obtained so far. Rather a large FS seems to be 
present at least at high temperature and the spectral weight 
around the anti-nodal regions is substantially suppressed with decreasing temperature, 
leading to Fermi arcs around the nodal region at low temperature \cite{norman98}. 
The origin of the Fermi arcs is directly related to the enigmatic phenomenon of the 
pseudogap, one of the most mysterious problems in high-$T_c$ cuprate superconductors \cite{timusk99}. 
The consistent understanding of quantum oscillation data and ARPES data is not 
obtained and the underlying FS in cuprates remains to be studied. 

ARPES measures the one-particle spectral function $A(\vk,\omega)$ and the underlying FS 
is usually obtained by tracing a peak position of  $A(\vk,\omega)$ at $\omega=0$. 
However, $A(\vk,\omega)$ features a broad structure especially 
near the anti-nodal region due to the pseudogap phenomenon \cite{yoshida12}. 
A technique complementary to ARPES is Compton scattering. 
It measures the momentum distribution function $n(\vk)$, which 
is obtained by integrating the product of $A(\vk,\omega)$ and 
the Fermi distribution function with respect to energy $\omega$. 
This feature can in turn become an advantage over ARPES, because 
$n(\vk)$ can be affected less severely by the opening of the gap around the anti-nodal region. 
In fact, the underlying FS can be inferred from $n(\vk)$ even in the presence of a gap due to 
superconductivity \cite{gyorffy89} and spin-density-wave \cite{friedel89}. 
Furthermore, in contrast to ARPES, no matrix-element effect occurs 
and $n(\vk)$ is imaged directly by Compton scattering in the whole Brillouin zone.  
As a result, we may reveal the underlying FS including the 
anti-nodal region by employing the Compton scattering technique. 
Compton scattering is actually a powerful probe to study fermiology. 
It is neither surface sensitive, in contrast to ARPES, nor disorder 
nor temperature sensitive unlike quantum oscillation measurements. 
Compton scattering successfully revealed the FS in La$_{2-x}$Sr$_{x}$CuO$_{4}$ 
(LSCO) with $x=0.3$ (Ref.~\onlinecite{al-sawai12}), Sr$_{2}$RuO$_{4}$ (Ref.~\onlinecite{hiraoka06}), 
CeRu$_{2}$Si$_{2}$ (Ref.~\onlinecite{koizumi11}), lithium \cite{tanaka01}, 
palladium-hydrogen \cite{mizusaki06}, and various disordered alloys such as 
Li-Mg  (Ref.~\onlinecite{stutz99}) and Cu-Pd (Ref.~\onlinecite{matsumoto01a}).

In this paper, we report high-resolution x-ray Compton scattering for LSCO with $x=0.08$. 
We find that the obtained momentum distribution function 
cannot be interpreted in terms of the conventional hole-like FS believed so far for La-based 
cuprates. A natural understanding is obtained by invoking 
the underlying electronic nematic correlations and their coupling to a soft phonon mode, 
which leads to a strongly deformed FS with $d$-wave symmetry in each 
CuO$_{2}$ plane and its alternate stacking along the $z$ axis. 
Additional Compton scattering measurements for $x=0.15$ and $0.30$ suggest 
similar FS deformation with weaker nematicity, which nearly vanishes at $x=0.30$. 

Our choice of $x=0.08$ is made judiciously. As is well known, LSCO 
exhibits both spin \cite{kimura99} and charge \cite{croft14} orders in $0.10 \lesssim x \lesssim 0.13$. 
This state is discussed in terms of spin-charge stripe order \cite{kivelson03}, where incommensurate 
magnetic order with wave vectors $\vq_{s}=(\pi,\pi\pm 2\pi\eta)$ and $(\pi\pm 2\pi\eta, \pi)$ coexists  
with one-dimensional charge stripe order with $\vq_{c} \approx \pm 2\vq_{s}$; 
we use the tetragonal notation in this paper. 
Moreover, the spin glass phase, diagonal spin-stripe phase, and antiferromagnetic phase 
are realized in a low temperature and low doping region \cite{niedermayer98,matsuda00}. 
Given that the present measurement is the first Compton scattering study to explore the underlying 
FS in the underdoped LSCO, apparent complications should be safely avoided. 
We therefore choose $x=0.08$ and a relatively high temperature region.  

\section{Results}
\begin{figure} [bht]
\centering
\includegraphics[width=16cm]{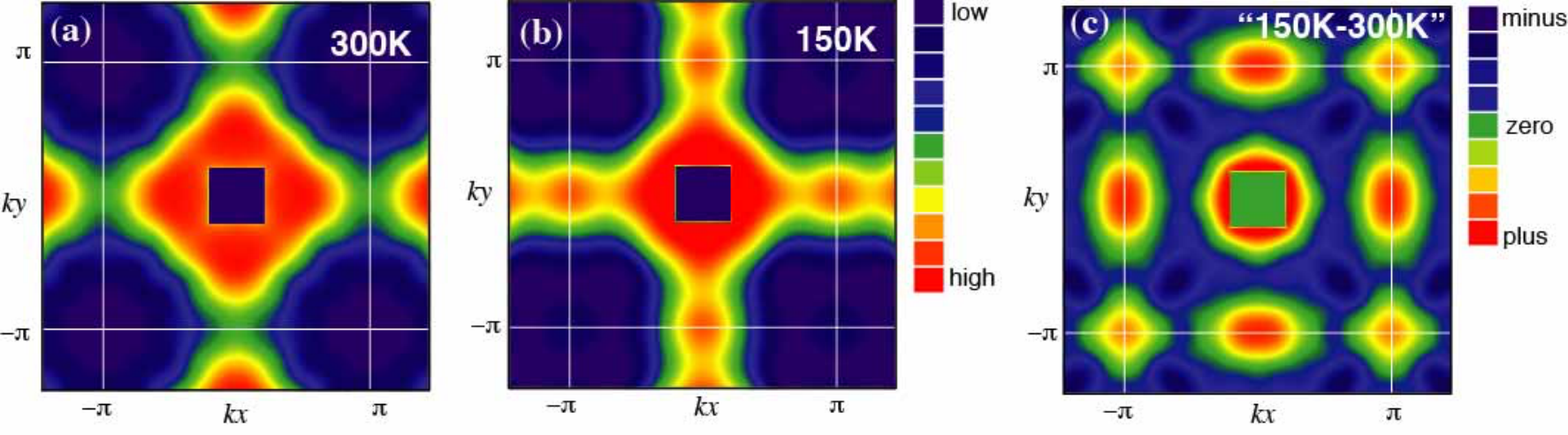}
\caption{(Color online) Images of the momentum distribution function $n(\vk)$ 
by high-resolution Compton scattering. 
Maps of $n(\vk)$ in the first Brillouin zone at $300$~K (a) and 150~K (b) 
for LSCO with $x=0.08$. 
The color scale represents the relative intensity. 
(c) The difference of $n(\vk)$ between $300$~K and 150~K. 
}  
\label{nk}
\end{figure}

We perform high-resolution Compton scattering for LSCO with $x=0.08$ at 300~K and 150~K. 
The momentum distribution function is then imaged in the 
two-dimensional Brillouin zone in Figs.~\ref{nk} (a) and (b) 
by applying the Lock-Crisp-West (LCW) theorem \cite{lock73}:  
higher intensity indicates a region where electrons are occupied more in momentum space. 
While all electrons contribute to the momentum distribution function in Compton scattering measurements, 
the momentum dependence of $n(\vk)$ originates from the band crossing the Fermi energy. 
Therefore $n(\vk)$ shown in Figs.~\ref{nk} (a)  and (b) reflects the underlying FS. 
The square region around $\vk=(0,0)$ contains a large experimental error and thus is not considered; 
see Experimental methods section for details. 

%%%%%%%%%%%%%%%%%%%%%% FIG. 1 %%%%%%%%%%%%%%%%%%%%%%%%
%%%%%%%%%%%%%%%%%%%%%%%%%%%%%%%%%%%%%%%%%%%%%%%%%

Usually the spectrum of $n(\vk)$ exhibits a weak temperature dependence even if a phase 
transition such as superconductivity occurs as long as translational symmetry is kept. 
However, as shown in Figs.~\ref{nk} (a) and (b),  
the present work reveals for the first time a strong temperature dependence of $n(\vk)$ 
for high-$T_c$ cuprate superconductors.  
Figure~\ref{nk} (c) is the difference of $n(\vk)$ between $300$~K and 150~K, 
showing that $n(\vk)$ is enhanced around $\vk=(\pi,0)$ and $(0,\pi)$ at 150~K. 
Since the pseudogap formation occurs below $200$~K (Ref.~\onlinecite{hashimoto07}), 
this change should be related to the pseudogap. 
However, the pseudogap is pronounced around $\vk=(\pi,0)$ and $(0,\pi)$, 
which would then suppress $n(\vk)$ there as long as the FS is hole-like, 
in contrast to the observation in \fig{nk}~(c). 
This apparent puzzle is solved by considering a combined effect of large FS deformation 
from the underlying nematicity and a broadening of $n(\vk)$ due to 
the pseudogap formation as we will analyze the data below. 
The enhancement around $\vk=(\pi,\pi)$ in \fig{nk}~(c) can be a consequence of 
the charge conservation.

For cuprates it is an open question how $n(\vk)$ should depend on $\vk$ both theoretically 
and experimentally; see Supplementary Material~A for a general feature of $n(\vk)$. 
Let us suppose a FS proposed by ARPES or tight-binding fitting and superpose it on our data of $n(\vk)$. 
In this case, one naturally expects that $n(\vk) > n(\vk_{F})$ [$n(\vk) < n(\vk_{F})$] 
for $\vk$ inside (outside) 
the FS and $n(\vk) \sim 0.5$ at $\vk=\vk_{F}$; 
here $\vk_{F}$ is the Fermi momentum. 
Therefore we reasonably assume that 
$n(\vk_{F})$ should not exhibit a strong $\vk_{F}$ dependence. 
However, this criterion itself may not be enough to discuss the underlying FS 
because the value of $n(\vk_F)$ cannot be 
exactly a certain value around 0.5. 
To compensate this uncertainty, we also study a peak position of the absolute value of 
the first derivative of $n(\vk)$, which should trace the underlying FS (Ref.~\onlinecite{hiraoka06}).  
In addition, we assume Luttinger's theorem that the volume enclosed by the FS 
is equal to the electron density. 
Note that our proposed FSs in Sec.~II~B also fulfills Luttinger's theorem.

\subsection{Conventional scenario} 
\begin{figure} [t]
\centering
\includegraphics[width=10cm]{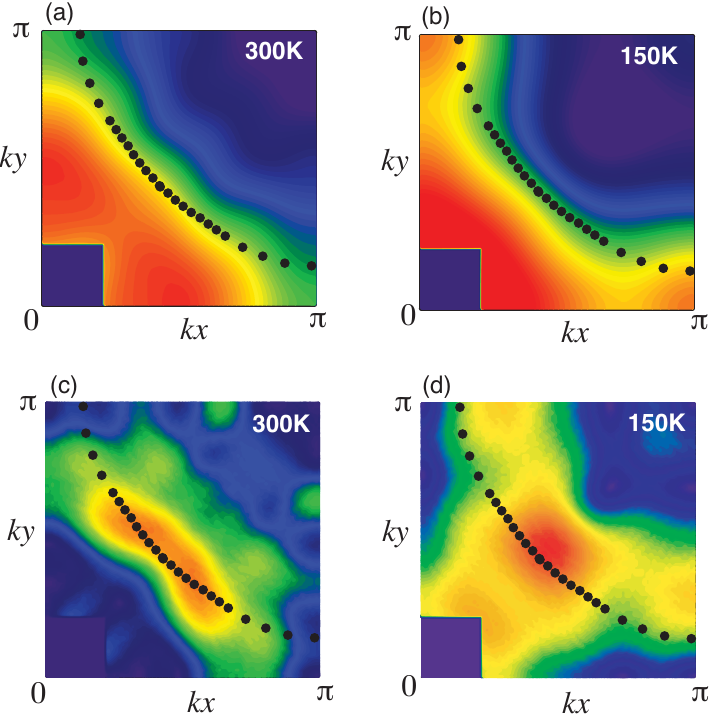}
\caption{(Color online)
Interpretation of Figs.~\ref{nk}~(a) and (b) in terms of the conventional FS (dots) 
reported by ARPES (Ref.~\onlinecite{yoshida12}). 
(a) and (c) are the momentum distribution function $n(\vk)$ and the magnitude of its 
derivative $|\nabla n(\vk)|$  at 300~K in the first quadrant of the Brillouin zone, respectively; 
(b) and (d) are the corresponding 
results at 150~K. 
While the ARPES measurements were performed for LSCO with 
$x=0.07$ (Ref.~\onlinecite{yoshida12}),  the doping difference 
by 1 \% from the present study provides no visible change of the FS 
on the scale of the figure. 
}  
\label{2DFS}
\end{figure}

We first consider the conventional hole-like FS, which is believed to be realized in 
La-based cuprates \cite{yoshida12}. As shown in Figs.~\ref{2DFS} (a) and (b), 
a comparison between our Compton data and the conventional FS reveals that 
$n(\vk_{F})$ exhibits a very weak $\vk_{F}$ dependence in an extended region 
around $\vk=(0.45\pi, 0.45\pi)$, consistent with the ARPES data \cite{yoshida12}. 
However, a sizable $\vk_{F}$ dependence is 
recognized in $\vk \approx (0.2\pi, 0.7\pi) - (0.1\pi,\pi)$ and its equivalent regions 
at both 300~K and 150~K. 
This is not consistent with a general understanding that 
$n(\vk_{F})$ should not exhibit  a strong $\vk_{F}$ dependence. 

%%%%%%%%%%%%%%%%%%%%%% FIG. 2 %%%%%%%%%%%%%%%%%%%%%%%%
%%%%%%%%%%%%%%%%%%%%%%%%%%%%%%%%%%%%%%%%%%%%%%%%%

Figures~\ref{2DFS} (c) and (d) show maps of the magnitude of the first derivative of $n(\vk)$, 
namely $| \nabla n(\vk) |$ at 300~K and 150~K, respectively, and the conventional FS is superposed there. 
At 300~K, the peak of $| \nabla n(\vk) |$ forms around $\vk=(0.45\pi, 0.45\pi)$ with an inward curvature, 
suggesting the presence of the electron-like FS around $\vk=(0,0)$. 
However, the conventional FS has the opposite curvature, not consistent with our data. 
At 150~K, the discrepancy between our data and the conventional FS is 
reduced. But an agreement is still not so satisfactory because 
the map of $| \nabla n(\vk) |$ has a peak around $(0.3\pi,\pi)$ and $(\pi, 0.3\pi)$, whose 
position is largely away from the conventional FS. 
We should not consider the weak peak structure around $(0.2\pi, 0.2\pi)$ in \fig{2DFS} (d), 
which can be an artifact coming from a large experimental error around $\vk=(0,0)$. 

Considering both $n(\vk)$ [Figs.~\ref{2DFS} (a) and (b)] and $| \nabla n(\vk) |$ [Figs.~\ref{2DFS} (c) and (d)], 
therefore, it is not possible to reconcile with the conventional FS both at 300~K and 150~K. 
To explore a possible clue to resolve such a problem, 
we first checked that the temperature dependence of the chemical potential is indeed negligible; 
see also Supplementary Material~C.  
While one might then wonder about the effect of the thermal broadening of $n(\vk)$, 
it is also not relevant to the present analysis; 
see Supplementary Material~D. 
Another idea to reconcile our data with the conventional FS (Ref.~\onlinecite{yoshida12}) 
would be to assume a strong 
temperature dependence of the band parameters such as the effective hopping integral $\tilde{t}^{\,'}$ 
between the next-nearest neighbor Cu sites. 
In this case, although a good agreement with $| \nabla n(\vk) |$ is not obtained, 
one could assume a small $| \tilde{t}^{\,'} |$ at 300~K, 
which then grows to be a large $| \tilde{t}^{\,'}|$ at 150~K and decreases  
at lower temperature, because the conventional FS is measured at 20~K (Ref.~\onlinecite{yoshida12}) 
and is fitted to a small $| \tilde{t}^{\,'} |$; see Supplementary Material~E for more details.  
However, it is not easy to explain such a strong temperature dependence of $\tilde{t}^{\,'}$. 

In addition, the pseudogap forms below $200$~K (Ref.~\onlinecite{hashimoto07}) and is most 
pronounced around $\vk=(\pi,0)$ and $(0,\pi)$. Since $\vk=(\pi,0)$ and $(0,\pi)$  are occupied by electrons 
for the conventional FS, $n(\vk)$ is expected to be suppressed there at 150 K because of a 
broadening of $n(\vk)$; 
see the inset in \fig{nk-1D-PG} and \fig{nk-1Dscan-2D} in Supplementary Material for explicit calculations. 
However, our data \fig{nk}~(c) shows the opposite, implying that the conventional FS 
is hard to be reconciled with our data.

\subsection{Nematic scenario} 
As an alternative scenario, we consider a possible effect of nematicity. 
Since nematicity is observed in other cuprates such 
as Y- \cite{ando02,hinkov08,cyr-choiniere15,sato17} and 
Bi--based \cite{nakata18,auvray19} cuprate compounds, 
it is natural to invoke the nematicity also in La-based cuprates. 
As the microscopic origin of the nematicity, several ideas are proposed: 
fluctuations of charge stripes \cite{kivelson98}, 
a $d$-wave Pomeranchuk instability \cite{yamase00a,yamase00b,metzner00}, 
an orbital order at oxygen sites \cite{fischer11,misc-oxygen}, 
and the nematicity from a biquadratic exchange interaction \cite{orth19}.

%%%%%%%%%%%%%%%%%%%%%% FIG. 3 %%%%%%%%%%%%%%%%%%%%%%%%
\begin{figure} [t]
\centering
\includegraphics[width=12cm]{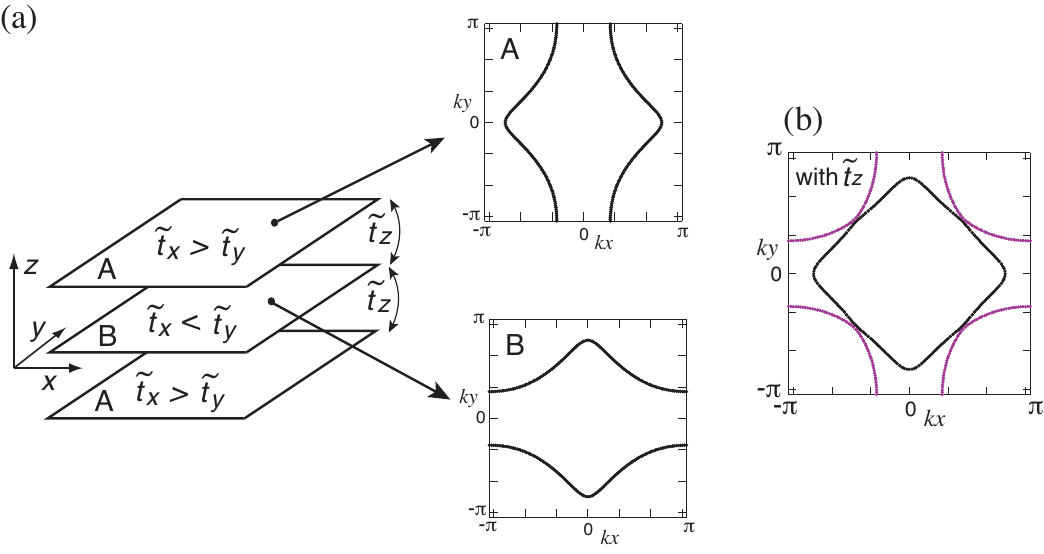}
\caption{(Color online) Fermi surface deformation in LSCO. 
(a) Alternate stacking of strongly deformed FSs due to the underlaying nematicity 
in each CuO$_{2}$ plane. $\tilde{t}_x$ and $\tilde{t}_y$ are 
the effective in-plane hopping integrals along the $x$ and $y$ directions, respectively; 
$\tilde{t}_z$ is the effective hopping along the $z$ direction; see Supplementary Material~B for more details. 
(b) The resulting bulk FSs, which  become two dimensional and consist of inner 
(black)  and outer (purple) FSs.  
}  
\label{stacking}
\end{figure}
%%%%%%%%%%%%%%%%%%%%%%%%%%%%%%%%%%%%%%%%%%%%%%%%%

LSCO with $x=0.08$ shows the so-called low-temperature orthorhombic (LTO) 
crystal structure below 350~K (Ref.~\onlinecite{birgeneau88}). 
This structure is characterized by the lattice anisotropy between [110] and [1$\bar{1}$0] direction, which 
does not couple to the nematic order. 
However, there is a low-energy phonon mode with energy less than 5 meV 
in the LTO structure \cite{kimura00a}. This phonon mode is referred to as the $Z$-point 
phonon mode and is related to the tilting mode of CuO$_{6}$ octahedra. It yields 
a {\it dynamical} anisotropy between [100] and [010] direction.
Consequently such a $xy$ anisotropy couples to the underlying nematic correlations 
and is strongly enhanced, leading to sizable nematicity. 
The resulting FS strongly deforms possibly to become 
a quasi-one-dimensional (Q1D) FS, i.e., an open FS. This deformation is 
expected to occur dynamically 
with an energy scale of the $Z$-point phonon mode. 
Since the $Z$-point phonon mode has anti-phase correlations along the $z$ axis, 
the Q1D FS is expected to stack alternately along the $z$ axis [see \fig{stacking}~(a)]. 
As a result, bulk FSs 
consist of two FSs: inner FS and outer FS as shown in \fig{stacking}~(b). 

%%%%%%%%%%%%%%%%%%%%%% FIG. 4 %%%%%%%%%%%%%%%%%%%%%%%%
\begin{figure} [t]
\centering
\includegraphics[width=10cm]{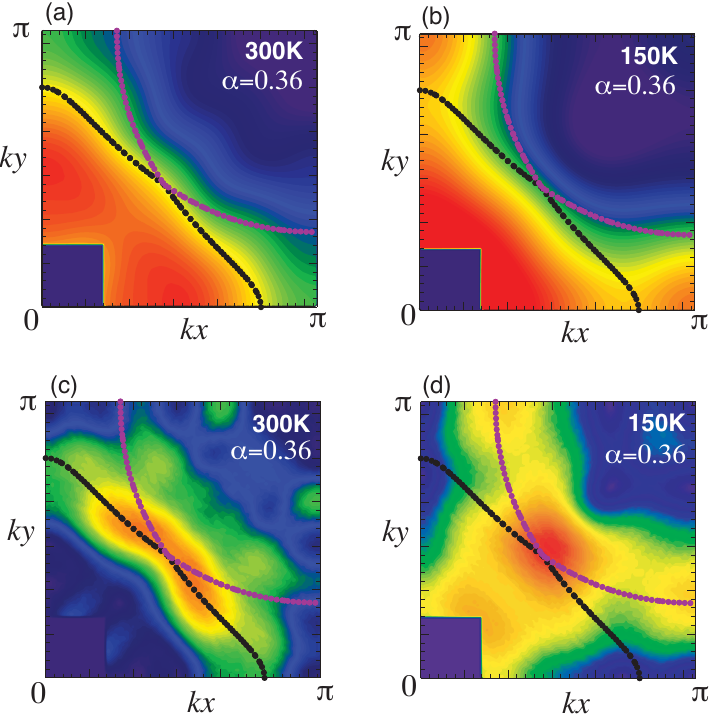}
\caption{(Color online)
Interpretation of Figs.~\ref{nk}~(a) and (b) in terms of the nematic scenario 
shown in \fig{stacking}. 
(a) and (c) are the momentum distribution function $n(\vk)$ and the magnitude of its 
derivative $|\nabla n(\vk)|$  at 300~K in the first quadrant of the Brillouin zone, respectively; 
(b) and (d) are the corresponding results at 150~K. 
A value of $\alpha$ controls the strength of nematicity and the conventional FS 
(Ref.~\onlinecite{yoshida12}) is reproduced at $\alpha=0$; 
see Supplementary Material~B for the precise definition of $\alpha$. 
}  
\label{nematic-FS}
\end{figure}
%%%%%%%%%%%%%%%%%%%%%%%%%%%%%%%%%%%%%%%%%%%%%%%%%

The energy scale of Compton scattering is 100 keV much larger than that 
of the $Z$-point phonon mode. The momentum distribution function observed by 
Compton scattering, therefore, reflects a snapshot of the alternate stacking 
of the Q1D FS, especially the most deformed shape where the "velocity" of the 
dynamical change of the FS becomes zero. 
This is indeed a reasonable approximation as we show in Supplementary Material~F. 

The idea shown in \fig{stacking}~(a)  was proposed a long time ago theoretically to interpret both 
magnetic excitation spectra and ARPES date consistently \cite{yamase00a,yamase01,yamase07} 
and now is supported by our Compton scattering data. 
While a Pomeranchuk instability was considered as a physics behind the strong nematicity 
in early studies \cite{yamase00a,yamase00b,yamase01,yamase07}, 
our experimental interpretation does not depend on the microscopic origin of the nematicity. 

Furthermore, our model \fig{stacking} is applicable to broader situations. 
First, the atomic pair distribution function analysis of neutron scattering data \cite{bozin99} 
suggests a possible local lattice distortion with $xy$-anisotropy, which alternates along the $z$-axis. 
This may yield the same stacking pattern as \fig{stacking} at least in a certain region around the local lattice distortion. 
Second, fluctuations associated with the $d$-wave Pomeranchuk instability can have anti-phase correlations 
between the layers  \cite{yamase09a}. In this case, our model \fig{stacking} is valid 
even without considering a coupling to other degrees of freedom. 
On the other hand, the model in \fig{stacking} cannot be applicable when there are no 
anti-phase nematic correlations along the $z$-axis. 
However, as long as nematic correlations are present in each CuO$_2$ plane,  
we expect that the resulting 
$n(\vk)$ can be interpreted in terms of the FSs shown in \fig{stacking}~(b). 
This is because the typical time scale of Compton scattering is much shorter 
than the electronic one and thus each Compton scattering event detects a snapshot of FS fluctuations 
and its statistical average is observed as a resulting image. 
When there are no nematic correlations, our model \fig{stacking} is reduced to a usual one where the conventional 
FS (Refs.~\onlinecite{yoshida12} and \onlinecite{yoshida06}) is realized in each CuO$_2$ plane.

\subsubsection{Underlying FSs}
On the basis of the nematic scenario, we therefore 
invoke the FSs in \fig{stacking}~(b) and superpose them on our data 
in Figs.~\ref{nematic-FS} (a)-(d). 
The inner FS at 300~K is located almost perfectly 
along the constant value of $n(\vk)$ [\fig{nematic-FS} (a)] 
and also reasonably traces along the peak position of the gradient of $n(\vk)$ [\fig{nematic-FS} (c)]. 
For the outer FS, 
the $\vk_{F}$ dependence of $n(\vk_{F})$ is reasonably small in \fig{nematic-FS} (a). 
Note that $\vk_{F}$ dependence of $n(\vk_{F})$ around $(0.3\pi, 0.9\pi)$ and $(0.9\pi, 0.3\pi)$ 
is actually small because the gradient of $n(\vk)$ there is very small as seen in \fig{nematic-FS} (c).   
The peak of $| \nabla n(\vk) |$ 
is well captured around $(0.45\pi,0.45\pi)$ in the red region in \fig{nematic-FS} (c). 
However, in contrast to the case of the inner FS, it does not extend along the outer FS 
and the gradient of $n(\vk)$ tends to be less pronounced 
when the momentum goes sufficiently away from $(0.45\pi, 0.45\pi)$.  
A possible reason is that the occupation number around the outer FS is smaller 
than that around the inner FS and is closer to the lowest occupation number 
[dark blue region in \fig{nematic-FS} (a)], which may make the gradient of $n(\vk)$ less pronounced 
compared with that around the inner FS.

At 150~K the outer FS is fully consistent with our data of $n(\vk)$ [\fig{nematic-FS} (b)] 
and is located at peak positions of $| \nabla n(\vk) |$  at 
$\vk=(0.3\pi,0)$, $(0,0.3\pi)$, and $(0.45\pi, 0.45\pi)$  [\fig{nematic-FS}~(d)]. 
For the inner FS, $n(\vk)$ is almost constant along the FS in an extended region 
around $(0.45\pi,0.45\pi)$, where $|\nabla n(\vk) |$ has a peak as expected. 
Since $| \nabla n(\vk) |$ becomes very small around $\vk=(0, 0.8\pi)$ and $(0.8\pi, 0)$, 
a seemingly sizable $\vk_{F}$ dependence of $n(\vk_{F})$ there in \fig{nematic-FS} (b) 
is actually not strong.

\subsubsection{Temperature dependence and the pseudogap effect} 
As shown in \fig{nk}~(c), the major difference between $300$~K and 150~K appears 
around $\vk=(\pi,0)$ and $(0,\pi)$ where $n(\vk)$ is {\it enhanced} at 150~K, 
although the pseudogap forms below $200$~K (Ref.~\onlinecite{hashimoto07}) and 
is pronounced at $\vk=(\pi,0)$ and $(0,\pi)$. 
This counterintuitive feature 
is a strong support of the nematic scenario shown in \fig{stacking}. 
To demonstrate this, we have performed a phenomenological analysis to model 
the pseudogap effect in terms of a strong damping of quasiparticles especially around 
$\vk=(\pi,0)$ and $(0,\pi)$; see Supplementary Material~G for details. 
Figure~\ref{nk-1D-PG} shows $n(\vk)$ calculated for several choices of damping $\Gamma_0$ 
along $(0,0)$-$(\pi,0)$-$(\pi,\pi)$ direction. 
Sharp drops at $\vk=(0.8\pi, 0)$ and $(\pi, 0.27\pi)$ correspond to the Fermi momenta.  
For $\Gamma_0=0$, $n(\vk)$ is already broad because of the effect of a finite temperature. 
With increasing the damping, $n(\vk)$ is broadened more, leading to an enhancement 
of $n(\vk)$ around $\vk=(\pi,0)$. This enhancement comes from the {\it asymmetric} broadening 
of $n(\vk)$ around the inner and outer FSs. 
Suppose the typical energy scale of the broadening is $\Gamma_{0}$, 
the corresponding broadening of momentum is then estimated as 
$\Delta \vk = \Delta E \frac{\Delta \vk }{\Delta E} \sim 2 \Gamma_{0} / {\bf v_{\vk}}$ 
with the velocity $\bf v_{\vk}$.  
For the typical dispersion in cuprates, $\bf v_{\vk}$ around $\vk=(0.8\pi,0)$ becomes 
much smaller than that around $(\pi,0.27\pi)$; 
see Eqs.~(\ref{vA}) and (\ref{vB}) in Supplementary Material~G.  
This momentum dependence  of $\bf v_{\vk}$ is the major reason of 
the substantial broadening of $n(\vk)$ around $\vk=(0.8\pi,0)$ with the damping of quasiparticles. 
Consequently the enhancement of $n(\vk)$ occurs around $\vk=(\pi,0)$.  

%%%%%%%%%%%%%%%%%%%%%% FIG. 5 %%%%%%%%%%%%%%%%%%%%%%%%
\begin{figure} [t]
\centering
\includegraphics[width=8cm]{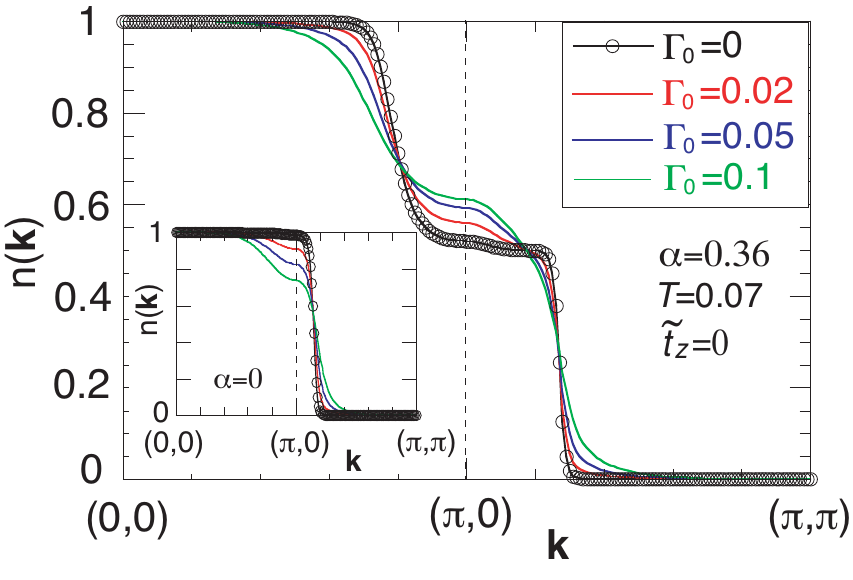}
\caption{(Color online)
Momentum distribution function $n(\vk)$ calculated in the presence of 
a damping of quasiparticles in the nematic scenario presented in \fig{stacking}. 
The damping depends on momentum, becomes the largest at $(\pi,0)$, 
and its magnitude is controlled by $\Gamma_{0}$; see Supplementary Material~G for details. 
The inset is the corresponding result for the conventional hole-like FS believed so far 
for La-based cuprates. 
}  
\label{nk-1D-PG}
\end{figure}
%%%%%%%%%%%%%%%%%%%%%%%%%%%%%%%%%%%%%%%%%%%%%%%%%

One may also invoke a gap formation especially 
around $\vk=(\pi,0)$ and $(0,\pi)$ to model the pseudogap phenomenology. 
In this case, $\Delta E$ corresponds to a gap value, instead of a damping $\Gamma_{0}$, 
and we still obtain results similar to \fig{nk-1D-PG}. 
The microscopic origin to yield  $\Delta E$, namely 
the origin of the pseudogap itself, remains elusive and 
is beyond the scope of the present measurements. 

At both 300~K and 150~K [Figs.~\ref{nematic-FS} (c) and (d)], 
the intensity becomes strong in an extended region around $(0.45\pi, 0.45\pi)$, 
where the inner and outer FSs are close to each other. This kind of feature is also observed in 
Sr$_{2}$RuO$_{4}$ (Ref.~\onlinecite{hiraoka06}), where three FSs nearly intersect around 
$(2\pi/3, 2\pi/3)$ and the intensity of the first derivative of $n(\vk)$ becomes strongest there in a wide region. 
The extended region of strong intensity at 300~K [\fig{nematic-FS}~(c)]
shrinks and concentrates around $\vk=(0.45\pi, 0.45\pi)$ at 150~K [\fig{nematic-FS}~(d)]. 
This is reasonable because the effect of the pseudogap is weak around $\vk=(0.45\pi, 0.45\pi)$ 
and the sharpest feature of $| \nabla n(\vk)|$ should be observed there. 

%%%%%%%%%%%%%%%%%%%%%% FIG. 6 %%%%%%%%%%%%%%%%%%%%%%%%
\begin{figure} [t]
\centering
\includegraphics[width=15cm]{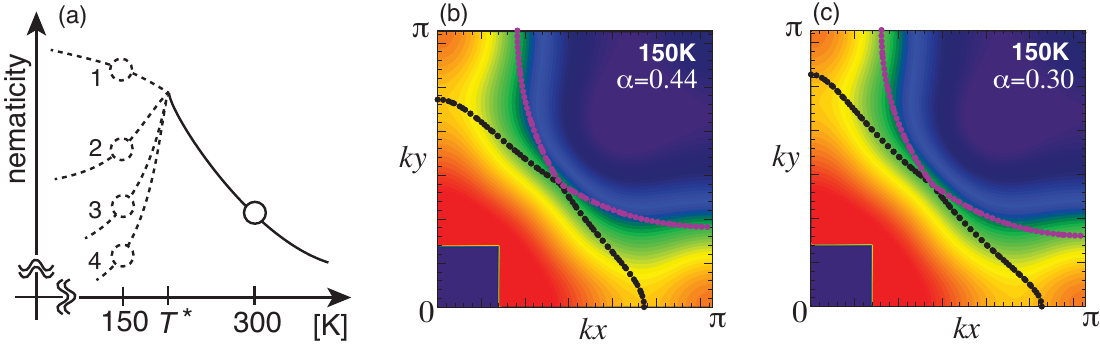}
\caption{(Color online) 
Possible temperature dependence of nematicity. 
(a) Sketch of temperature dependence of the nematicity. The pseudogap temperature $T^{*}$ 
is estimated around 200~K (Ref.~\onlinecite{hashimoto07}). 
Given that the system is close to a nematic instability, the nematicity is expected to increase 
with decreasing temperature at least down to $T^{*}$. 
It is not clear how the nematicity develops below $T^{*}$ 
and thus four possible scenarios are shown: the nematicity increases (scenario 1) 
or decreases (scenarios 2-4).   In the latter case, compared with the nematicity at 300~K, 
the nematicity at 150~K becomes larger (scenario 2), comparable (scenario 3), or smaller (scenario 4). 
(b) and (c) are typical FSs for the scenarios 1 and 2, and scenario 4, respectively; 
they are superposed on the observed map of $n(\vk)$ at 150~K. 
The value of $\alpha$ is a parameter of the nematicity and 
the conventional FS (Ref.~\onlinecite{yoshida12}) is reproduced for $\alpha=0$. 
See Supplementary Material~H for a comparison with the gradient of $n(\vk)$. 
}  
\label{T-depend}
\end{figure}
%%%%%%%%%%%%%%%%%%%%%%%%%%%%%%%%%%%%%%%%%%%%%%%%%

Because the large anisotropy comes from the underlying nematic correlations in the electron 
system, the anisotropy is expected to have a strong temperature dependence 
as indicated theoretically \cite{yamase00b,orth19}. 
Since the so-called pseudogap temperature $T^{*}$ is around 200~K for $x=0.08$ (Ref.~\onlinecite{hashimoto07}) 
and the gap-like feature develops around $\vk=(\pi,0)$ and $(0,\pi)$ below $T^{*}$, 
the temperature dependence of the nematicity likely changes below $T^{*}$. 
As sketched in \fig{T-depend}~(a),  
there are two possibilities below $T^{*}$: the nematicity still develops or is suppressed. 
In the  latter case, it is not clear whether the nematicity at 150~K becomes larger 
or smaller than that at 300~K or comparable to that. 
Figures~\ref{nematic-FS}~(b), \ref{T-depend}~(b) and \ref{T-depend}~(c) indicate that 
our data at 150~K is equally well fitted by FSs with different degrees of the nematicity; 
see Supplementary Material~H for more details. 
This ambiguity comes from a rather broad feature of $n(\vk)$ and its derivative, 
which is mainly due to strong electron correlations typical to cuprate superconductors.  
It remains to be studied which scenario shown in \fig{T-depend}~(a) is likely appropriate to LSCO. 
In addition, the determination of the "onset" temperature of the nematicity 
is also left to a future study.

\section{Concluding remarks} 
The present Compton scattering experiments provide three important insights into 
the electronic property in cuprate superconductors. 

First, our observed images for La-based cuprates imply that the FS is strongly deformed by the 
underlying nematicity in each CuO$_2$ plane, but the bulk FSs recover the fourfold symmetry. 
The resulting electronic property does not exhibit anisotropy in bulk in spite of the strong nematicity. 
This is in total contrast to Y-based cuprates, where  the intrinsic small 
$xy$-anisotropy is strongly enhanced 
as observed by resistivity \cite{ando02}, neutron scattering \cite{hinkov08}, 
Nernst coefficient \cite{cyr-choiniere15}, and magnetic torque \cite{sato17}. 

Second, the momentum distribution function is enhanced around $\vk=(\pi,0)$ and $(0,\pi)$ 
in the pseudogap phase, although the pseudogap itself is most pronounced there. 
This counterintuitive feature implies the presence of the reconstructed FSs due to the underlying nematiciy.

Third, our observed images cannot be reconciled with the conventional hole-like FS (Ref.~\onlinecite{yoshida12}). 
Nonetheless, our proposed FSs are not inconsistent with the ARPES spectra. 
ARPES is most precise around the so-called nodal region for cuprates. 
In fact,  the position of our FSs near $\vk=(0.45\pi,0.45\pi)$ almost 
coincides with the FS from ARPES (Ref.~\onlinecite{yoshida12}); compare \fig{nematic-FS} with \fig{2DFS}. 
Near $\vk=(0,\pi)$, on the other hand, the ARPES data in Refs.~\onlinecite{yoshida06,he09} 
show a broad peak, which was interpreted as a single peak. 
Such a broad peak may also be interpreted as consisting of 
the underlying double-peak structure originating from the inner and outer FSs as implied from 
the present Compton scattering experiments. 
The complementary employment of ARPES and Compton scattering will provide more detailed 
information to elucidate the electronic structure in cuprates. 

We have focused on the underdoped LSCO with $x=0.08$. 
It is natural to ask how the nematicity and the shape of the FSs evolve with increasing doping. 
At a fixed temperature, nematic correlations are expected to be less pronounced with increasing 
doping rate and a conventional FS may be realized eventually in the heavily overdoped region. 
By performing Compton scattering measurements for LSCO with $x=0.15$ and $0.30$ 
at 300~K, we have confirmed that 
the data at $x=0.15$ can be understood with weaker nematicity than $x=0.08$, 
whereas those at $x=0.30$ imply 
a conventional electron-like FS but possibly tiny nematicity cannot be excluded. 
Details are given in Supplementary Material~K.

FS deformation associated with nematicity is also observed 
in Fe-based superconductors \cite{nakayama14}. 
Its effect is, however, not so drastic as the one that we have found in La-based cuprates. 
This is because the FS in cuprates is located close to $\vk=(\pi,0)$ and $(0,\pi)$, and thus 
the nematicity can change easily the FS topology.

Finally, the present experiments indicate that Compton scattering can be a powerful tool 
to elucidate the FS and work beyond the widely-employed techniques such as ARPES and 
quantum oscillations. 
Since Compton scattering is neither disorder nor surface sensitive and is available 
also in the presence of electric and magnetic fields, it will be exciting 
that Compton scattering is employed as a complementary tool to 
ARPES and quantum oscillations in various correlated electron systems.

\section{Experimental methods} 
High-resolution Compton scattering requires a large single crystal to map the 
momentum distribution function. Our high-quality single crystals of 
LSCO with $x=0.08$, 0.15, and 0.30 were grown by a traveling-solvent floating-zone method.  
The feed rod was prepared by the solid state reaction from La$_2$O$_3$, SrCO$_3$, and CuO 
in the molar ratio of $(2-x)/2: x: 1.0$. The mixed powder was sintered at $\sim1000$ $\degree$C 
for 24 hours in air with intermediate grinding. The sintered powder was then pressed into 
the cylindrical rod and again 
sintered at $1250$~$\degree$C for 24 hours in air. 
We used this feed rod as well as a solvent with a composition of La$_{2-x}$Sr$_{x}$CuO$_4$ : CuO$_2$ = $1: 4.7$ and La$_2$CuO$_4$ as a seed rod for 
the crystal growth in an infrared radiation furnace with oxygen-gas flow rate 
of 100 cm$^3$/min. The growth rate was set to be 1.0 mm/h. 
We obtained a 100 mm-long crystal rod and annealed it in oxygen gas flow 
to minimize oxygen deficiencies. A part of the grown rod was cut into the cubic-shaped portion about 
the size of 5$\times$5$\times$5 mm$^3$ for our Compton measurements. 
We measured the magnetic susceptibility, and confirmed the superconducting transition 
with $T_c$ $\approx$ 20~K, 38~K, and 0~K at $x=0.08$, 0.15, and 0.30, respectively, from the Meissner signal. 

We measured Compton profiles with scattering vectors equally spaced between 
the [100] and [110] directions using the Cauchois-type x-ray spectrometer at the BL08W beamline 
of SPring-8. 
The overall momentum resolution is estimated to be 0.13 a.u. in FWHM. 
The incident x-ray energy was 115 keV and the scattering angle was 165 degrees. 
The Compton-scattered x-rays were measured by a two-dimensional position-sensitive detector. 
Energy distribution of the Compton-scattered x-rays was centered at about 80 keV. 
Approximately 3 $\times$ 10$^5$ counts in total were collected at the Compton peak channel. 
Each Compton profile was corrected for absorption, analyzer, detector efficiencies, 
scattering cross section, possible double scattering contributions, and x-ray background \cite{cooper04}. 
A two-dimensional momentum density, representing a projection of the three-dimensional momentum 
density onto the $ab$ plane (see Supplementary Material~I for the effect of $k_z$ dispersion), 
was reconstructed from each set of ten Compton profiles 
by using the direct Fourier transform method \cite{matsumoto01}.
This method produces unphysical oscillations at low momenta, which is the reason why 
there occur large errors in the central area of the experimental $n(\vk)$. 
To fold the obtained $n(\vk)$ into a single central Brillouin zone, we employ 
the LCW theorem \cite{lock73}.  This theorem is exact for non-interacting electrons, 
but an approximation for interacting electrons.

%\newpage

\vspace{5mm}
{\bf Data availability} 
The data that support the finding of this study are available from H.Y. and Y.S. upon reasonable  request.

\acknowledgments
The authors thank M. Ito for technical supports of Compton scattering measurements and  
A. Fujimori, W. Metzner, 
M. Ogata, and H. Yokoyama for valuable discussions on the interpretation 
of the present Compton scattering data. 
H.Y. was supported by a Grant-in-Aid for Scientific Research Grant Number JP15K05189, 
JP18K18744, JP20H01856, and by JST-Mirai Program Grant Number JPMJMI18A3, Japan. 

\vspace{5mm}
{\bf Author contributions}

H.Y. initiated this study and Y.S. performed the Compton scattering measurements. 
H.Y. and Y.S. analyzed the data together and H.Y. wrote the major part of the manuscript 
with the help of K.Y. and Y.S.   
High-quality single crystals were prepared by M.F., S.W., and K.Y. 

\bibliography{main} 

\begin{thebibliography}{10}
\bibitem{misc-oxygen}
\bibinfo{note}{The orbital order at oxygen sites is essentially the same as the
  $d$-wave Pomeranchuk instability, which is obtained as a bond order with
  wavevector zero and leads to an intra-unit cell order at oxygen sites with
  wavevector zero.}
\bibitem{misc-kf}
\bibinfo{note}{Here $\vk > \vk_{F}$ and $\vk < \vk_{F}$ should be understood as
  the momentum $\vk$ outside and inside the Fermi surface, respectively.}
\end{thebibliography}

\newpage

%\appendix
\section*{\Large Supplementary Material}
\subsection{General features of $\boldsymbol{n(\vk)}$} 
\begin{figure} [t]
\centering
\includegraphics[width=10cm]{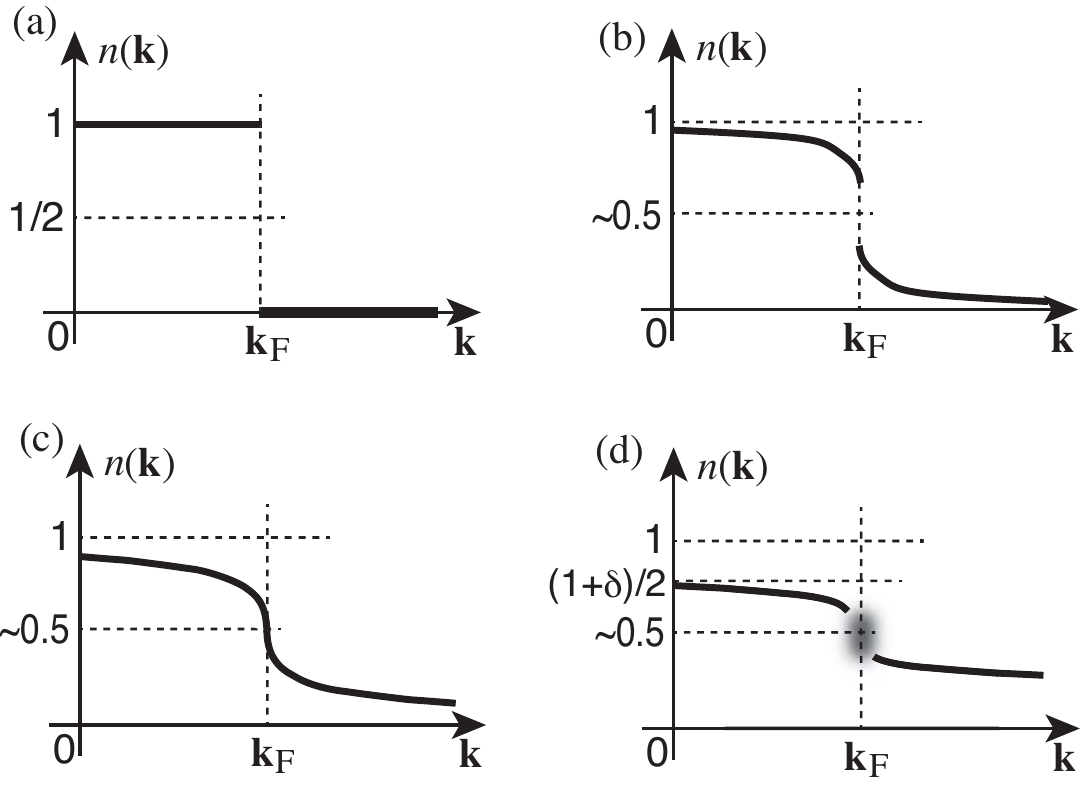}
\caption{(Color online)
Sketches of the momentum distribution function at zero temperature in typical cases: 
(a) non-interaction system, (b) Fermi liquid system, (c) non-Fermi liquid system such as 
a Tomonaga-Luttinger liquid, 
and (d) the limit of the strong correlation. It is not clear whether $n(\vk)$ exhibits a jump 
or a smooth evolution around $\vk_{F}$ in (d).  
}  
\label{nk-explain}
\end{figure}

The momentum distribution function $n(\vk)$ is a central quantity in the present work. 
Here we recall its general feature. 
In a noninteracting electron system [\fig{nk-explain}~(a)], $n(\vk)=1$ in $\vk < \vk_{F}$, 
$0.5$ at $\vk=\vk_F$, and 0 in $\vk < \vk_F$ (Ref.~\onlinecite{misc-kf}); 
here $\vk_{F}$ is Fermi momentum. 
In a Fermi liquid [\fig{nk-explain}~(b)], 
$n(\vk)$ is similar to the noninteracting case, but the magnitude of 
a jump of $n(\vk)$ at $\vk=\vk_{F}$ 
is reduced to $Z_{F} < 1$. Still $n(\vk_{F})$ is expected {\it around} 0.5; 
it is not necessarily to be exactly 0.5. 
In a non-Fermi liquid state [\fig{nk-explain}~(c)] such as 
a Tomonaga-Luttinger liquid well established in a one-dimensional system, 
$n(\vk)$ exhibits power-law behavior around $\vk_{F}$ and there is no 
jump at $\vk_{F}$.  A value of $n(\vk_{F})$ is usually around 0.5. 
In the limit of strong correlations where double occupancy of electrons is forbidden [\fig{nk-explain}~(d)],  
a general insight can be obtained. 
Suppose hole carriers are introduced into the half-filled 
antiferromagnetic Mott insulator. When the carrier density is $\delta$, we obtain 
the sum rule \cite{stephan91} 
\be
\int_{-\infty}^{\infty} A(\vk,\omega) {\rm d}\omega = \frac{1}{2} (1+\delta)\,.
\label{sum-rule}
\ee
Hence it follows that for any $\vk$ point 
\be
n(\vk) = \int_{-\infty}^{\infty} A(\vk,\omega) f(\omega) {\rm d}\omega \leq \frac{1}{2} (1+\delta)\,.
\label{nk-strong}
\ee
Equations~(\ref{sum-rule}) and (\ref{nk-strong}) are exact relations. 
At $\delta=0$ where one electron resides at each site, the electron density is given by 
$\frac{2}{N} \sum_{\vk} n(\vk) = 1$, where $N$ is the total number of lattice sites 
and the factor of 2 counts the spin degrees of freedom. 
Equation~(\ref{nk-strong}) then implies $n(\vk)=0.5$ independent of $\vk$ at $\delta=0$. 
When carriers are doped and the system becomes metallic, we have the FS. 
We then  expect 
$n(\vk) > n(\vk_{F})$ and $n(\vk) < n(\vk_{F})$ for $\vk$ inside and outside the FS, respectively, 
and $n(\vk_{F}) \sim 0.5$ on the FS,  
as indeed found in a variational Monte Carlo study in the $t$-$J$ model \cite{pruschke91,sato18}.  

%%%%%%%%%%%%%%%%%%%%%% FIG. 7 %%%%%%%%%%%%%%%%%%%%%%%%
%%%%%%%%%%%%%%%%%%%%%%%%%%%%%%%%%%%%%%%%%%%%%%%%%

\subsection{Formalism of momentum distribution function}
It is insightful to see how the momentum distribution function looks in the simplest case for a 
realistic layered system of La-based cuprates where the unit cell 
contains two different layers and each layer shifts by $[\frac{1}{2},\frac{1}{2},\frac{1}{2}]$ 
(Ref.~\onlinecite{radaelli94}). We denote each layer by the $A$ and $B$ plane as shown in 
\fig{stacking}~(a) and we consider the following minimal one-body Hamiltonian: 
\be
\tilde{\mathcal{H}} = \sum_{\vk \sigma}  
\left( 
\begin{array}{cc}
c_{\vk \sigma}^{A \dagger}  & c_{\vk \sigma}^{B \dagger}
\end{array}
\right) 
\left(
\begin{array}{cc}
\xi_{\vk}^{A} & \epsilon_{\vk} \\
\epsilon_{\vk} & \xi_{\vk}^{B}
\end{array}
\right)
\left(
\begin{array}{c}
c_{\vk \sigma}^{A} \\
c_{\vk \sigma}^{B}
\end{array} 
\right) \,.
\label{effective-model}
\ee
Here $c_{\vk \sigma}^{A \dagger} (c_{\vk \sigma}^{A})$ is the creation (annihilation) operator of 
electrons with momentum $\vk$ and spin $\sigma$ in the $A$ plane, $\xi_{\vk}^{A}$ the in-plane 
dispersion, and $\epsilon_{\vk}$ the $c$-axis dispersion. 
Similarly the corresponding quantities are defined for the $B$ plane. This one-body Hamiltonian 
can be regarded as an effective Hamiltonian and may be obtained after a mean-field approximation 
to an interacting electron system. 
It is straightforward to compute the eigenenergies  of the Hamiltonian: 
\be
\lambda_{\pm}(\vk) = \frac{\xi_{\vk}^{A} + \xi_{\vk}^{B} }{2} \pm \sqrt{\left( 
\frac{\xi_{\vk}^{A} - \xi_{\vk}^{B} }{2} \right)^{2} + \epsilon_{\vk}^{2}}\,. 
\label{eigen}
\ee

We normalize the momentum distribution function $n(\vk)$ in a way that its maximal value is unity: 
\be
n(\vk) = \frac{1}{4} \sum_{\sigma} \left( \left\bra  c_{\vk \sigma}^{A \dagger}c_{\vk \sigma}^{A}  \right\ket 
+  \left\bra  c_{\vk \sigma}^{B \dagger}c_{\vk \sigma}^{B}  \right\ket  \right) \,.
\ee 
Using the Fermi distribution function $f(\lambda) = 1/({\rm e}^{\lambda/T}+1)$, where $T$ is temperature, 
the right-hand side of the above equation is easily computed as 
\bea 
n(\vk) &&= \frac{1}{2}\int_{-\infty}^{\infty} d\omega \left( A_{+}(\vk,\omega) + A_{-}(\vk,\omega)\right)  
f(\omega)   \,, \label{nk-spectral} \\
&& = \frac{1}{2} \left( 1- \frac{1}{2} \tanh \frac{\lambda_{+}(\vk)}{2T} 
- \frac{1}{2} \tanh \frac{\lambda_{-}(\vk)}{2T} \right) 
\label{nk-eigen} \,.
\eea
Here we have introduced the spectral function $A_{\pm}(\vk,\omega)$ for a later convenience, 
which is given by $A_{\pm}(\vk,\omega)=\delta(\omega -  \lambda_{\pm}(\vk))$ in the present case. 
When $\xi_{\vk} =\xi_{\vk}^{A} = \xi_{\vk}^{B}$ and $\epsilon_{\vk}=0$, the above 
expression is reduced to the well-known Fermi distribution function 
\be
n(\vk) = \frac{1}{2} \left( 1- \tanh \frac{\xi_{\vk}}{2T} \right) \, .
\ee

Since the CuO$_{2}$ plane forms a square lattice and we will allow $xy$ anisotropy of the band dispersion,  
we parameterize the in-plane dispersion as 
\be
\xi_{\vk}^{A} = -2 \left(  \tilde{t}_x \cos k_x + \tilde{t}_y \cos k_y \right ) 
- 4 \tilde{t}^{\,'} \cos k_x \cos k_y 
- 2  \left(  \tilde{t}_x^{\,''} \cos 2k_x + \tilde{t}_y^{\,''} \cos 2k_y \right )  - \mu 
\label{xia}
\ee
where $\tilde{t}_x = \tilde{t} ( 1+ \frac{1}{2} \alpha)$,  $\tilde{t}_y = \tilde{t} ( 1- \frac{1}{2} \alpha)$, 
$\tilde{t}_x^{\,''} = \tilde{t}^{\,''} ( 1+ \frac{1}{2} \alpha'')$, 
$\tilde{t}_y^{\,''} = \tilde{t}^{\,''} ( 1- \frac{1}{2} \alpha'')$, and 
$\mu$ is the chemical potential. 
Because of possible coupling to the $Z$-point phonons \cite{kimura00a}, 
the in-plane dispersion in the adjacent plane may have the opposite anisotropy, namely 
\be
\xi_{\vk}^{B} = -2 \left(  \tilde{t}_y \cos k_x + \tilde{t}_x \cos k_y \right ) 
- 4 \tilde{t}^{\,'} \cos k_x \cos k_y 
- 2  \left(  \tilde{t}_y^{\,''} \cos 2k_x + \tilde{t}_x^{\,''} \cos 2k_y \right )  - \mu \,.
\label{xib}
\ee 
The $A$ and $B$ planes are shifted by $[\frac{1}{2},\frac{1}{2},\frac{1}{2}]$ with each other. 
The $c$-axis dispersion may be parameterized by 
\be
\epsilon_{\vk} = - 8 \tilde{t}_z \cos \frac{k_x}{2} \cos \frac{k_y}{2} \cos \frac{k_z}{2} \,. 
\label{c-axis}
\ee
The ARPES study \cite{horio18} suggests a different form 
\be
\epsilon_{\vk} = - 2 \tilde{t}_z \cos \frac{k_x}{2} \cos \frac{k_y}{2} \cos \frac{k_z}{2} 
\left[ (\cos k_x - \cos k_y)^{2} + c_0 \right]\,,
\label{c-axis2}
\ee
with $c_0=0$. This functional form shall also be considered.  

We first take $\alpha=\alpha''=0$ and $\tilde{t}_z = 0$ and fit the FS to the ARPES data 
at $x=0.07$ (Ref.~\onlinecite{yoshida12}) under the condition $\tilde{t}^{\,''} = -\tilde{t}^{\,'}/2$ 
(Ref.~\onlinecite{andersen95}), which yields the following values: 
\be
\tilde{t}^{\,'}=-0.15 \tilde{t}, \quad \tilde{t}^{\,''} = 0.075 \tilde{t} \quad {\rm for}\,\, x=0.07\,.
\label{tp-value}
\ee 
At $x=0.22$,  $\tilde{t}_{z}$ is estimated around $0.07 \tilde{t}$ in the analysis 
of the ARPES data at $12$~K  (Ref.~\onlinecite{horio18}). 
The coherency along the $z$ direction is best achieved in the superconducting state and 
likely becomes worse with increasing temperature. 
In fact, the $c$-axis resistivity is non-metallic in the state above $T_c$ (Ref.~\onlinecite{ito91}). 
Hence in our measurements at 300~K and 150~K 
a value of  $\tilde{t}_{z}$ may be smaller than $\tilde{t}_{z}= 0.07 \tilde{t}$. 
In addition, because of the strong correlation effect specific to 
cuprate superconductors, the effective value of $\tilde{t}_{z}$ is proportional 
to the carrier density as indeed the case of the $t$-$J$ model \cite{yamase01}. 
This effect also works to suppress a value of  $\tilde{t}_{z}$ at $x=0.08$. 
Given a likely small value of $\tilde{t}_{z}$ (see the subsection~J for possible values of $\tilde{t}_z$), 
it turns out that our conclusions obtained 
in the present work do not depend on a precise choice of $\tilde{t}_{z}$ nor on 
a choice of the $c$-axis dispersions [Eqs.~(\ref{c-axis}) and (\ref{c-axis2})].   
Hence to keep our presentation as simple as possible, 
we took $\tilde{t}_{z}=0$ to compute the FSs in Figs.~\ref{nematic-FS} and \ref{T-depend}.

We measure all quantities with the dimension of energy in units of $\tilde{t}$. 
Our parameters $\alpha$ and $\alpha''$ determine a band anisotropy 
and we assume  $\alpha = \alpha''$ for simplicity. The chemical potential $\mu$ is determined 
to reproduce the doping rate. 
$n(\vk)$ depends on not only in-plane momentum but also out-of-plane momentum. 
Since our Compton scattering data integrate $k_z$ dependence, we consider 
\be
n(k_x, k_y) = \frac{1}{N_z} \sum_{k_z} n(\vk)  
\label{nk-average}
\ee
where $N_z$ is a half of the number of the layers, namely the number of the unit cell along the $z$ direction. 
We took $N_z=48$, which is sufficiently large. To keep the same notation as the main text 
we write $n(k_x, k_y)$ as $n(\vk)$ below.

\subsection{Fermi surfaces} 
The FS is obtained by solving $\lambda_{\pm} (\vk) =0$ in \eq{eigen} 
and always fulfills Luttinger's theorem. 
Since the unit cell contains two layers, we generally obtain two FSs for each $k_z$. 
In principle, the FS depends on temperature via the chemical potential $\mu$. 
However, such an effect is negligible in a temperature range that we are interested in. 
Rather, as we discussed in \fig{T-depend}~(a), 
the sizable temperature dependence is expected  for the parameter 
$\alpha$ introduced in the band dispersions [see Eqs.~(\ref{xia}) and (\ref{xib})] 
due to the underlying nematic correlations. 

\subsection{Thermal broadening effect} 
\begin{figure} [t]
\centering
\includegraphics[width=8cm]{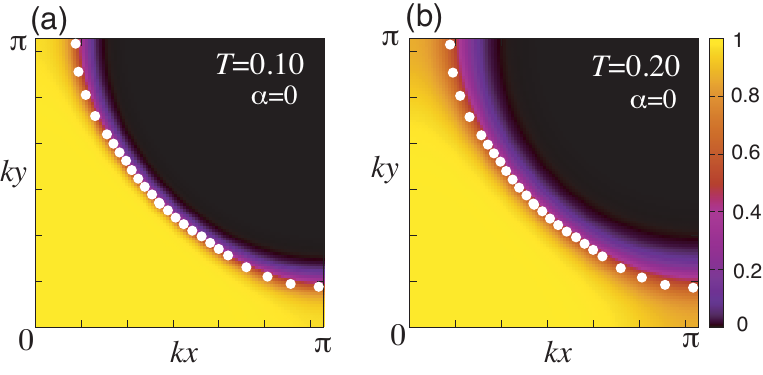}
\caption{(Color online) 
Momentum distribution function $n(\vk)$ of the Hamiltonian (\ref{effective-model}) 
at $T=0.1$ (a) and $0.2$ (b). The band parameters are chosen to reproduce the FS 
reported in the ARPES measurements (white circles) \cite{yoshida12}; see \eq{tp-value}. 
}  
\label{2Dnk-T}
\end{figure}

To see the thermal broadening effect on $n(\vk)$, we compute $n(\vk)$  for $\alpha =0$ 
and $\tilde{t_{z}}=0$.  
Figure~\ref{2Dnk-T} shows the results in the first quadrant of the Brillouin zone 
at $T=0.1$ and $0.2$, which may be associated with the data at 150~K and 300~ K, respectively. 
We superpose the FS reported by ARPES (Ref.~\onlinecite{yoshida12}) on \fig{2Dnk-T}.  
Since electron correlation effects are considered only via the effective one-body Hamiltonian 
in the present calculations [see \eq{effective-model}], 
$n(\vk)$ shows a sharp drop at the Fermi momentum $\vk_{F}$. 
With increasing $T$, $n(\vk)$ is broadened around $\vk_{F}$, but 
we see no signature that $n(\vk_{F})$ starts to depend on $\vk_{F}$ by thermal broadening. 

It is apparent that as long as we stick to the conventional FS, the map of $n(\vk)$ shown in 
\fig{2Dnk-T} is hard to be reconciled with our data [Figs.~\ref{2DFS} (a) and (b)]. 
In fact, a comparison with \fig{2DFS} implies 
a possible internal structure in the yellow region in \fig{2Dnk-T}. 
Our idea of the nematicity actually produces an inner FS as we discussed in \fig{stacking}. 

%%%%%%%%%%%%%%%%%%%%%% FIG. 8 %%%%%%%%%%%%%%%%%%%%%%%%
%%%%%%%%%%%%%%%%%%%%%%%%%%%%%%%%%%%%%%%%%%%%%%%%%

\subsection{Possible cure of the conventional Fermi surface} 
\begin{figure} [t]
\centering
\includegraphics[width=8cm]{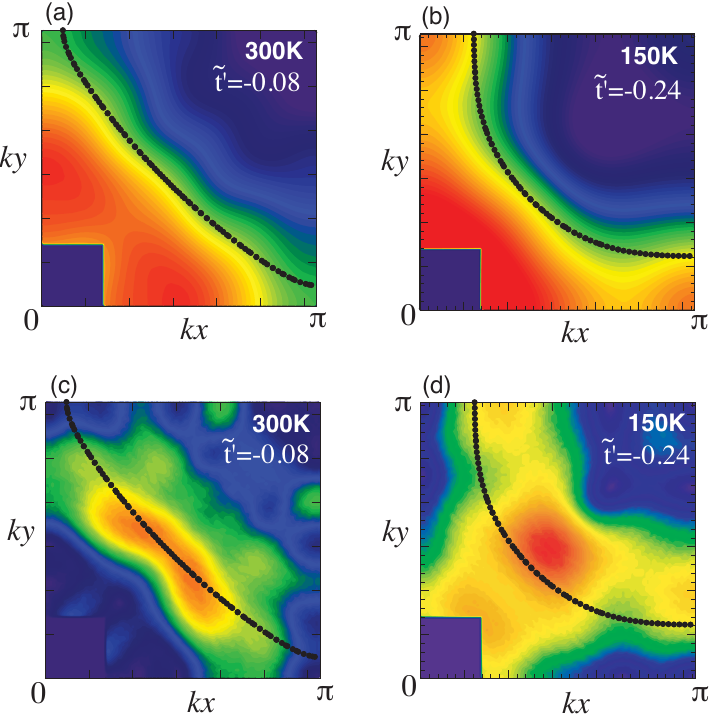}
\caption{(Color online)
Fit to our data (\fig{nk}) in terms of a conventional FS, namely 
without nematicity ($\alpha=0$) at 300~K (a) and 150~K (b). The band parameter $\tilde{t}^{\,'}$ 
is taken as $\tilde{t}^{\,'}=-0.08$ and $-0.24$ at 300~K and 150~K, respectively. 
(c) and (d) the corresponding maps of the derivative of $n(\vk)$. 
}  
\label{2D-fit}
\end{figure}

It is difficult to reconcile the conventional FS reported by ARPES (Ref.~\onlinecite{yoshida12}) 
with our data (\fig{2DFS}) as we discussed in Sec. II A as well as the above subsection~D. 
A possible cure of this conventional idea may be to introduce a temperature dependence 
of the band parameter $\tilde{t}^{\,'}$ under the condition $\tilde{t}^{\,''} = -\tilde{t}^{\,'}/2$ 
(Ref.~\onlinecite{andersen95}). 
We would then assume around $\tilde{t}^{\,'}=-0.08$ 
at 300~K and $\tilde{t}^{\,'}=-0.24$ at 150~K so that 
$n(\vk_{F})$ has a weak $\vk_{F}$ dependence [Figs.~\ref{2D-fit} (a) and (b)], 
although the agreement with the map of $| \nabla n(\vk) |$ is not so satisfactory 
in the sense that the curvature around $\vk=(0.45\pi, 0.45\pi)$ is different at 300~K 
[\fig{2D-fit} (c)] and 
the conventional FS is away from the peak position of $| \nabla n(\vk) |$ 
at $(0.3\pi, \pi)$ and $(\pi, 0.3\pi)$ at 150~K [\fig{2D-fit} (d)]. 
In addition, the FS seems to deviate sufficiently from the strongest peak of $| \nabla n(\vk) |$ 
around $\vk=(0.45\pi, 0.45\pi)$  at 150~K in \fig{2D-fit} (d).
Recalling that ARPES data \cite{yoshida12} was taken at 20~K and 
the reported FS is well fitted with $\tilde{t}^{\,'}=-0.15$ there 
when we stick to the conventional FS [see \eq{tp-value}],  
we would need to assume a strong and non-monotonous temperature dependence of $\tilde{t}^{\,'}$ 
between 300~K and 20~K, which does not seem realistic. 

%%%%%%%%%%%%%%%%%%%%%% FIG. 9 %%%%%%%%%%%%%%%%%%%%%
%%%%%%%%%%%%%%%%%%%%%%%%%%%%%%%%%%%%%%%%%%%%%%%%%

\subsection{Time averaging effect of FS fluctuations} 
\begin{figure} [t]
\centering
\includegraphics[width=4cm]{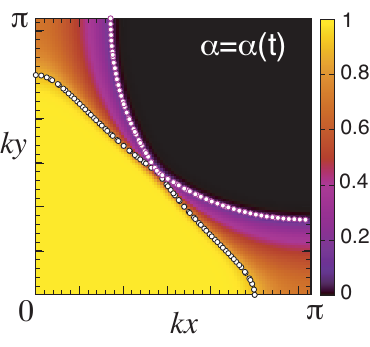}
\caption{(Color online)
Time-averaged momentum distribution function $\bar{n}(\vk)$ for $\alpha_{\rm max}=0.36$ 
[Eqs.~(\ref{alpha-t}) and (\ref{T-average-nk})] at $T=0.07$ and $\tilde{t}_{z}=0$. 
The FS for  $\alpha=\alpha_{\rm max}$ is superposed (circles). 
}  
\label{time-average} 
\end{figure}

We consider a coupling to the $Z$-point phonon mode \cite{kimura00a}. 
In this case, the anisotropy of the FS is expected to be fluctuating in time with the same scale 
as the phonon, which is less than 5 meV. 
On the other hand, Compton scattering is a high-energy probe and observes a snapshot 
of the fluctuating FS at each time. The resulting signal is time-averaged. 
We model the time-dependent anisotropy of the FS as 
\be
\alpha(t)=\alpha_{\rm max} \cos(\omega_{Z} t + \phi)
\label{alpha-t}
\ee
where $\omega_{Z}$ is the frequency of the $Z$-point phonons, $t$ time, and $\phi$ a phase. 
We may write the momentum distribution function $n(\vk)$ as $n(\vk, \alpha(t))$. 
The time-averaged momentum distribution function $\bar{n}(\vk)$ is then given as 
\be
\bar{n}(\vk) = \frac{1}{T_Z} \int_{0}^{T_Z} n(\vk, \alpha(t)) {\rm d}t \,,
\label{T-average-nk}
\ee 
where $T_Z=2\pi/\omega_{Z}$ is the periodicity in time. 
$\bar{n}(\vk)$ for $\alpha_{\rm max}=0.36$ is shown in \fig{time-average}. 
While $\bar{n}(\vk)$ exhibits a broadened feature, it is well characterized by the FS with 
$\alpha=\alpha_{\rm max}$. 

%%%%%%%%%%%%%%%%%%%%%% FIG. 10 %%%%%%%%%%%%%%%%%%%%%
%%%%%%%%%%%%%%%%%%%%%%%%%%%%%%%%%%%%%%%%%%%%%%%%%

\subsection{Phenomenological study of a pseudogap effect} 
Here we present a full description of a phenomenological study of a pseudogap effect on $n(\vk)$ 
in terms of a quasiparticle damping; 
the essential part is already given in the main text. 

As a minimal model, we replace the spectral function in \eq{nk-spectral} as 
\be
A_{\pm}(\vk,\omega) = \frac{1}{\pi} \frac{\Gamma_{\vk}}{(\omega - \lambda_{\pm}(\vk))^2 + \Gamma_{\vk}^2} \,,
\label{spectral-A}
\ee
where $\Gamma_{\vk}$ represents a damping of quasiparticles. Since the pseudogap effect 
is most pronounced 
around $\vk=(\pi,0)$ and $(0,\pi)$ (Ref.~\onlinecite{timusk99}), 
we assume the following $\vk$ dependence: 
\be
\Gamma_{\vk} = \Gamma_{0} (\cos k_x - \cos k_y)^{2} + T^{2} \,.
\label{damping}
\ee
The second term is a regular contribution and relevant only around the nodal direction where 
the first term vanishes. We compute $n(\vk)$ under the condition of charge conservation, 
namely $n= \frac{1}{N} \sum_{\vk \sigma} n(\vk) = 1-\delta$, where $N$ is the total number of the lattice cites. 
We neglect the effect of $c$-axis dispersion, which is irrelevant to the 
present analysis, 

%%%%%%%%%%%%%%%%%%%%%% FIG. 11 %%%%%%%%%%%%%%%%%%%%%
\begin{figure} [t]
\centering
\includegraphics[width=8cm]{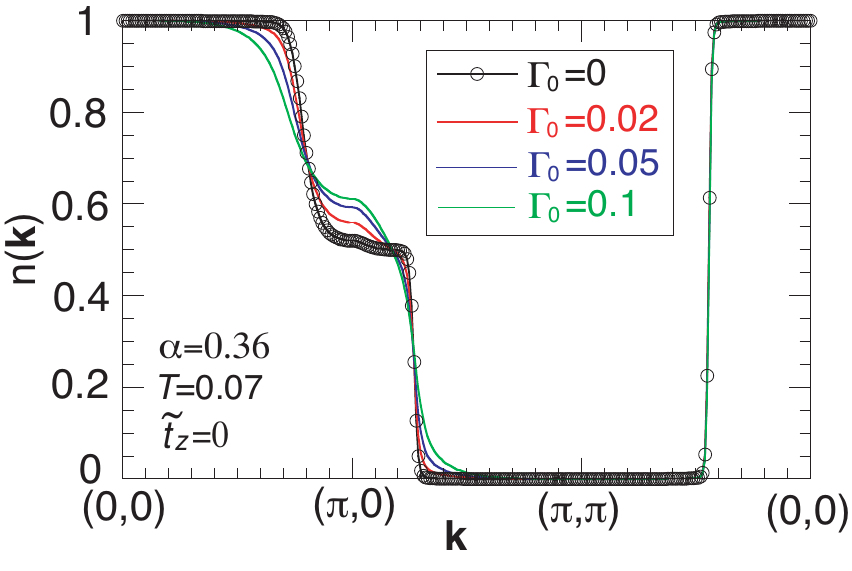}
\caption{(Color online) 
Momentum distribution function $n(\vk)$ of the Hamiltonian (\ref{effective-model}) 
for $\alpha=0.36$ at $T=0.07$ and $\delta=0.08$. 
Strength of quasiparticle damping is parameterized by $\Gamma_{0}$ in \eq{damping}. 
}  
\label{nk-1Dscan}
\end{figure}
%%%%%%%%%%%%%%%%%%%%%%%%%%%%%%%%%%%%%%%%%%%%%%%%%

In \fig{nk-1Dscan} we plot $n(\vk)$ along the symmetry axis for several choices of $\Gamma_{0}$ at $T=0.07$. 
$n(\vk)$ shows rapid changes at $\vk=(0.8\pi,0)$, $(\pi,0.27\pi)$, and $(0.45\pi, 0,45\pi)$, which 
correspond to the Fermi momenta for $\alpha=0.36$. 
The sharp feature around  $(0.45\pi, 0,45\pi)$ stays even with increasing $\Gamma_{0}$, 
because the $d$-wave form factor in \eq{damping} vanishes. 
On the other hand, $n(\vk)$ is broadened with $\Gamma_{0}$ 
around $(0.8\pi,0)$ and $(\pi,0.27\pi)$ and the broadening is more pronounced around  $(0.8\pi,0)$. 
This {\it asymmetry} originates from the typical band structure of cuprates. 
The quasiparticle dispersions [\eq{eigen}] are given 
by $\xi_{\vk}^{A}$ [\eq{xia}] and  $\xi_{\vk}^{B}$  [\eq{xib}] when we neglect the $c$-axis dispersion. 
$\xi_{\vk}^{A}$ crosses the FS around $\vk=(0.8\pi,0)$, 
whereas  $\xi_{\vk}^{B}$ does around $\vk=(\pi,0.27\pi)$. 
The velocity of $\xi_{\vk}^{A}$ at $\vk=(k_x,0)$ is given by 
\be
{\bf v}_{\vk}^{A}= \left( 2(\tilde{t}_x + 2 \tilde{t}^{\,'} + 4\tilde{t}_x^{\,''} \cos k_x) \sin k_x, 0 \right) \,,
\label{vA}
\ee
and that of $\xi_{\vk}^{B}$ along $\vk = (\pi, k_y)$ is 
\be
{\bf v}_{\vk}^{B}= \left(0,  2(\tilde{t}_x - 2 \tilde{t}^{\,'} + 4\tilde{t}_x^{\,''} \cos k_y) \sin k_y \right) \,. 
\label{vB}
\ee
We obtain $|{\bf v}_{\vk}^{A}| = 0.69$ at $\vk=(0.8\pi,0)$ and $|{\bf v}_{\vk}^{B} |= 2.57$ at $\vk=(\pi,0.27\pi)$ 
for the present parameters. This big difference comes from the presence of $\tilde{t}^{\,'}$ and 
$\tilde{t}^{\,''}$. 
Since the quasiparticle damping $\Gamma_{\vk}$ gives rise to the 
broadening of momentum as 
$\Delta \vk = \Delta E \frac{\Delta \vk}{\Delta E} \sim 2 \Gamma_{\vk} / {\bf v_{\vk}}$, 
the momentum distribution function $n(\vk)$ becomes much broader around 
$\vk= (0.8\pi,0)$ than $(\pi, 0.27\pi)$.  In addition, 
compared  with the result for $\Gamma_{0}=0$, 
this broadening gives rise to an {\it increase} of the occupation around $\vk=(\pi,0)$, 
although the damping $\Gamma_{\vk}$ is biggest there. 
This unexpected feature comes from the presence of two Fermi momenta around $\vk=(\pi,0)$, 
typical to the nematic scenario (\fig{stacking}).

Since the pseudogap forms below 200~K and its maximal energy scale is around 40 meV  (Ref.~\onlinecite{hashimoto07}), 
we may associate the results for $\Gamma_{0} \approx 0.05 - 0.1(\approx 0 - 0.01)$ to our data 
at 150~K (300~K), considering $\tilde{t} \sim 100$ meV. 
This phenomenological analysis explains 
i) $n(\vk)$ is broadened around $\vk=(0.8\pi,0)$ substantially more than 
around $\vk = (\pi, 0.27\pi)$ at 150~K 
as observed in Figs.~\ref{nematic-FS}~(b) and (d),  
ii) the resulting signal is enhanced around $\vk=(\pi,0)$ and $(0,\pi)$ compared with the data at 300~K  
[see Fig.~\ref{nk}~(c)], and 
iii) the region around $\vk=(0.45\pi, 0.45\pi)$ stays essentially the same at both 300~K and 150~K 
[see Figs.~\ref{nematic-FS} (c) and (d)]. 

%%%%%%%%%%%%%%%%%%%%%% FIG. 12 %%%%%%%%%%%%%%%%%%%%%
\begin{figure} [t]
\centering
\includegraphics[width=8cm]{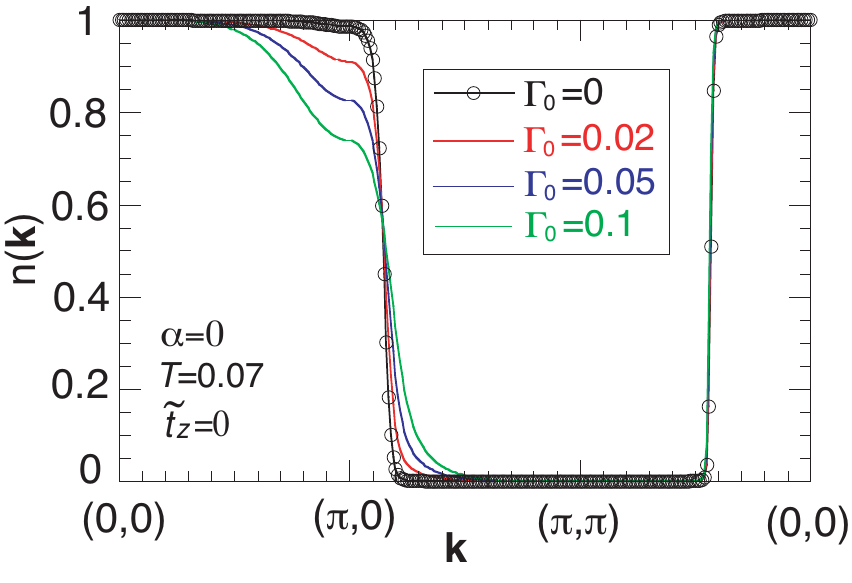}
\caption{(Color online) 
Momentum distribution function $n(\vk)$ of the Hamiltonian (\ref{effective-model}) 
for $\alpha=0$ at $T=0.07$ and $\delta=0.08$. 
Strength of quasiparticle damping is parameterized by $\Gamma_{0}$ in \eq{damping}. 
}  
\label{nk-1Dscan-2D}
\end{figure}
%%%%%%%%%%%%%%%%%%%%%%%%%%%%%%%%%%%%%%%%%%%%%%%%%

In the present simple analysis,  
we have introduced the damping of the Lorentzian form [\eq{spectral-A}] into the effective 
Hamiltonian (\ref{effective-model}). Hence the maximum and minimum values of $n(\vk)$ become 
1 and 0, respectively. Physically, however, in cuprates we would expect the maximum value 
of $n(\vk)$ is around $(1+\delta)/2$ [see \eq{nk-strong}] and the minimum value is well above zero as seen 
in variational Monte Carlo study in the $t$-$J$ model \cite{sato18}. 
In addition, the position of our FS for a finite $\Gamma_{0}$ around $\vk=(0.8\pi,0)$ 
slightly shifts to a smaller momentum in \fig{nk-1Dscan}, because of the shift of the chemical 
potential due to the presence of $\Gamma_{\vk}$. 
Although this shift comes from the charge conservation and is a similar mechanism of the 
shift of the chemical potential with temperature, our obtained shift might be an artifact of the simplicity of 
the present analysis, which neglects incoherent contributions to the spectral function $A_{\pm} (\vk,\omega)$  
[see \eq{spectral-A}]. In this sense, the present analysis should be regarded as
a demonstration of the asymmetry of the broadening in $n(\vk)$ around $\vk=(\pi,0)$ 
and of the enhancement of $n(\vk)$ there. These come from the underlying band structure 
near the Fermi energy and thus are expected to be robust features. 
 While we have introduced the damping of quasiparticles $\Gamma_{\vk}$ to 
capture the pseudogap phenomenology, a result similar to \fig{nk-1Dscan} is also 
obtained by invoking a gap in the electronic dispersion when its energy scale is 
comparable to our $\Gamma_{\vk}$.

We also present in \fig{nk-1Dscan-2D} the corresponding results for $\alpha=0$,  
namely for the conventional FS reported by ARPES (Ref.~\onlinecite{yoshida12}). 
$n(\vk)$ crosses the FS at $\vk=(\pi,0.14\pi)$ and $(0.43\pi,0.43\pi)$. 
The broadening due to the damping of quasiparticles is visible around $\vk=(\pi,0)$ and 
is pronounced along the $(0,0)$-$(\pi,0)$ direction more than the $(\pi,0)$-$(\pi,\pi)$ direction, 
because the velocity along the $(0,0)$-$(\pi,0)$ direction is smaller 
as we have explained in Eqs.~(\ref{vA}) and (\ref{vB}). 
Since $\vk=(\pi,0)$ is located inside the FS, $n(\vk)$ is suppressed there due to the damping 
of quasiparticles, in sharp contrast to the case in \fig{nk-1Dscan}.

\subsection{Degree of nematicity} 
\begin{figure} [ht]
\centering
\includegraphics[width=8cm]{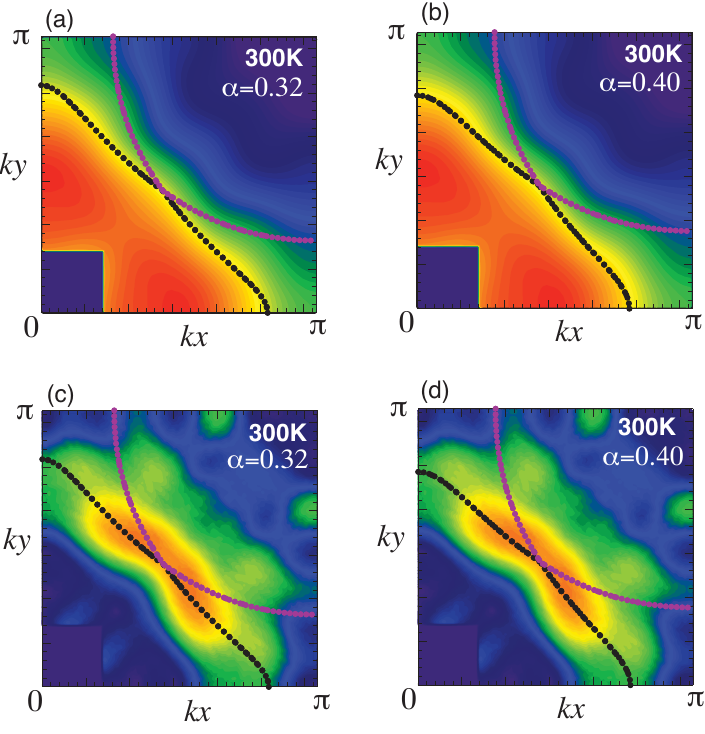}
\caption{(Color online) 
Interpretation of the observed momentum distribution function $n(\vk)$ at 300~K with 
a different degree of nematicity: $\alpha=0.32$ (a) and $0.40$ (b). 
(c) and (d) the corresponding data of the derivative of  $n(\vk)$.  
}  
\label{nematic-T300}
\end{figure}

\begin{figure} [t]
\centering
\includegraphics[width=8cm]{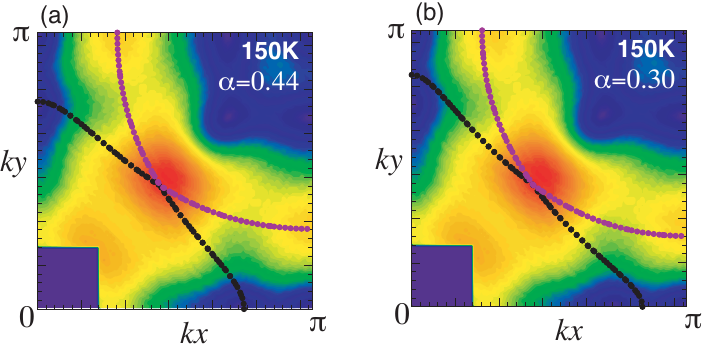}
\caption{(Color online) 
Interpretation of the observed maps of the derivative of 
momentum distribution function $n(\vk)$ at 150~K with 
a different degree of nematicity: $\alpha=0.44$ (a) and $0.30$ (b).  
The corresponding maps of $n(\vk)$ were shown in Figs.~\ref{T-depend} (b) and (c). 
}  
\label{nematic-T150}
\end{figure}

As seen in Eqs.~(\ref{xia}) and (\ref{xib}), nematicity is parameterized by $\alpha$ in our formalism. 
We have taken $\alpha=0.36$ in \fig{nematic-FS} to be consistent with our data. 
The degree of the nematicity, however, cannot be determined uniquely from the present 
data because of a rather broad feature of $n(\vk)$ and its derivative. 
In fact, an equally good fit is obtained in $0.32 \lesssim \alpha \lesssim 0.40$ at 300~K 
as shown in \fig{nematic-T300}. The situation is also the same at $150$~K and 
we may invoke a value of $\alpha$ at least in $0.30< \alpha < 0.44$, as was 
demonstrated in Figs.~\ref{T-depend} (b) and (c). The corresponding data of $| \nabla n(\vk)  |$ 
is shown in Figs.~\ref{nematic-T150} (a) and (b).  

%%%%%%%%%%%%%%%%%%%%%% FIG. 13 %%%%%%%%%%%%%%%%%%%%%
%%%%%%%%%%%%%%%%%%%%%%%%%%%%%%%%%%%%%%%%%%%%%%%%%

%%%%%%%%%%%%%%%%%%%%%% FIG. 14 %%%%%%%%%%%%%%%%%%%%%
%%%%%%%%%%%%%%%%%%%%%%%%%%%%%%%%%%%%%%%%%%%%%%%%%

\subsection{Effect of $\boldsymbol{k_z}$ dispersion} 
\begin{figure} [t]
\centering
\includegraphics[width=8cm]{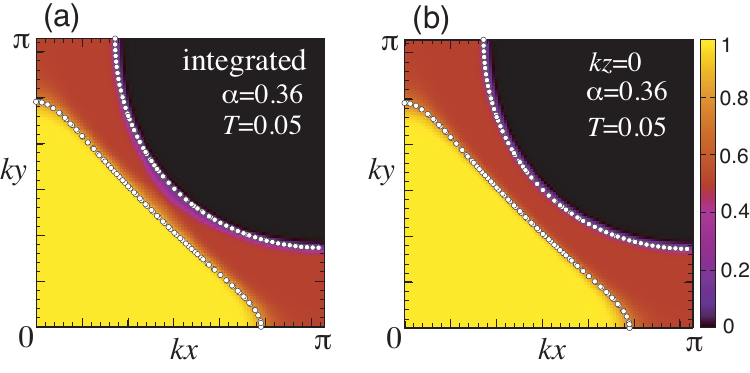}
\caption{(Color online) 
Momentum distribution function $n(\vk)$ of the Hamiltonian (\ref{effective-model}) 
for $\alpha=0.36$ 
and $\tilde{t}_{z}=0.1$ at $T=0.05$ and $\delta=0.08$:  
(a) $k_z$ dependence is integrated [see \eq{nk-average}] and (b) $k_z=0$ is taken. 
The FSs obtained for $k_z=0$ are superposed on both figures. 
}  
\label{kz-effect-nk}
\end{figure}

In the actual Compton scattering measurements, the spectrum 
is integrated along the $k_z$ direction [\eq{nk-average}]. 
It is therefore insightful to clarify the effect of the $k_z$ integration. Figure~\ref{kz-effect-nk} compares 
$n(\vk)$ for $k_z=0$ with that after integration with respect to $k_z$; 
the FSs for $k_z=0$ are also superposed there.  
It is clear that the effect of $k_z$ integration is very weak 
and $n(\vk)$ in \fig{kz-effect-nk} (a) is well captured in terms of the FSs for $k_z=0$ 
even if we take $\tilde{t}_{z}$ larger than the realistic value.

%%%%%%%%%%%%%%%%%%%%%% FIG. 15 %%%%%%%%%%%%%%%%%%%%%
%%%%%%%%%%%%%%%%%%%%%%%%%%%%%%%%%%%%%%%%%%%%%%%%%

\subsection{Magnitude of $\boldsymbol{c}$-axis dispersion} 
%%%%%%%%%%%%%%%%%%%%%% FIG. 16 %%%%%%%%%%%%%%%%%%%%%
\begin{figure} [hb]
\centering
\includegraphics[width=8cm]{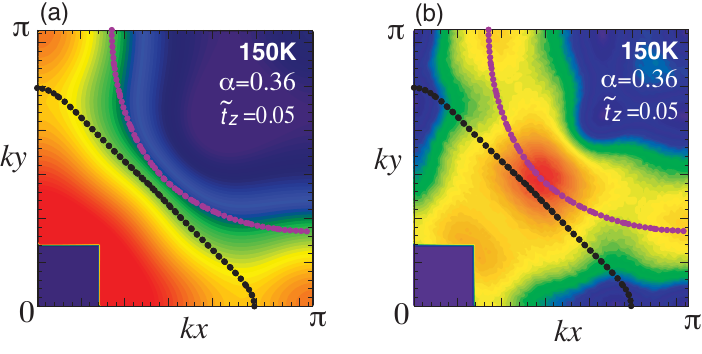}
\caption{(Color online) 
Interpretation of the observed maps of the momentum distribution function $n(\vk)$ (a) 
and its derivative $| \nabla n(\vk) |$ (b) at 150~K 
in terms of the FSs with $\alpha=0.36$ and $k_z=0$ 
for $\tilde{t}_{z}=0.05$ in \eq{c-axis}; the corresponding FS for $\tilde{t}_{z}=0$ is shown in 
Figs.~\ref{nematic-FS} (b) and (d). 
}  
\label{tz-fit}
\end{figure}
%%%%%%%%%%%%%%%%%%%%%%%%%%%%%%%%%%%%%%%%%%%%%%%%%
As discussed in the subsection~B, the actual value of $\tilde{t}_{z}$ is not known for LSCO 
with $x=0.08$. 
As long as $\tilde{t}_{z}$ is sufficiently small, a difference between the 
$c$-axis dispersions [Eqs.~(\ref{c-axis}) and (\ref{c-axis2})] is minor. 
On the basis of a comparison between our proposed FS (\fig{stacking}) and the observed 
momentum distribution function presented in the main text,  
we may estimate $\tilde{t}_{z} \lesssim 0.01$ at 300~K. 
On the other hand, at 150~K it is possible to invoke a relatively large value such as 
$\tilde{t}_{z} \approx 0.05$ 
for the $c$-axis dispersion \eq{c-axis}. The resulting FSs for $k_z=0$ are superposed on \fig{tz-fit}. 
Compared with Figs.~\ref{nematic-FS}~(b) and (d), where we took $\tilde{t}_{z}=0$, 
the major difference appears around $\vk=(0.45\pi, 0.45\pi)$ due to the sizable interlayer coupling. 
Still the outer FS almost perfectly agrees with our data and in addition, 
the $\vk_{F}$ dependence of $n(\vk_{F})$ is also weak entirely along the inner FS. 
We can conclude that the FSs with $\tilde{t}_{z}=0.05$ are also consistent with our data. 
While the $c$-axis dispersion \eq{c-axis2} with $c_0=0$ vanishes around $\vk=(0.45\pi, 0.45\pi)$ 
because of the factor of $(\cos k_x - \cos k_y)^{2}$, 
a split between the outer and inner FSs there is easily obtained by considering 
disorder effects which disrupt the symmetry of the bonding orbital \cite{xiang96}. 
In this case, one may invoke a finite value of $c_0$ in \eq{c-axis2}.

\newpage

\subsection{Compton scattering for $\boldsymbol{x=0.15}$ and 0.30}
We have focused on the doping rate $x=0.08$ so far and showed that the FS can be strongly deformed 
by the underlying nematicity, but the bulk FSs recover the fourfold symmetry. 
It is natural to ask how the FS deformation evolves with increasing doping. 
At a fixed temperature, nematic correlations are expected to be less pronounced with increasing doping and 
eventually a conventional FS suggested by ARPES may be realized in heavily overdoped LSCO such as $x=0.30$. 
This tendency also collaborates on the Z-point phonon mode, 
which is present in $x \leq 0.21$ (Refs.~\onlinecite{birgeneau88} and \onlinecite{kimura00a}).  

In this section, we shall present data that the nematicity indeed becomes smaller at $x=0.15$ and almost vanishes at $x=0.30$ 
by performing additional Compton scattering measurements. To make a consistent comparison with the data at $x=0.08$, 
we choose a temperature 300~K instead of 150~K. This is because a signature of short-range charge order is reported above 
the pseudogap temperature at $x=0.15$ (Ref.~\onlinecite{jjwen19}) and we can safely avoid potential complications 
from that.

%%%%%%%%%%%%%%%%%%%%%% FIG. 17 %%%%%%%%%%%%%%%%%%%%%
\begin{figure} [t]
\centering
\includegraphics[width=12cm]{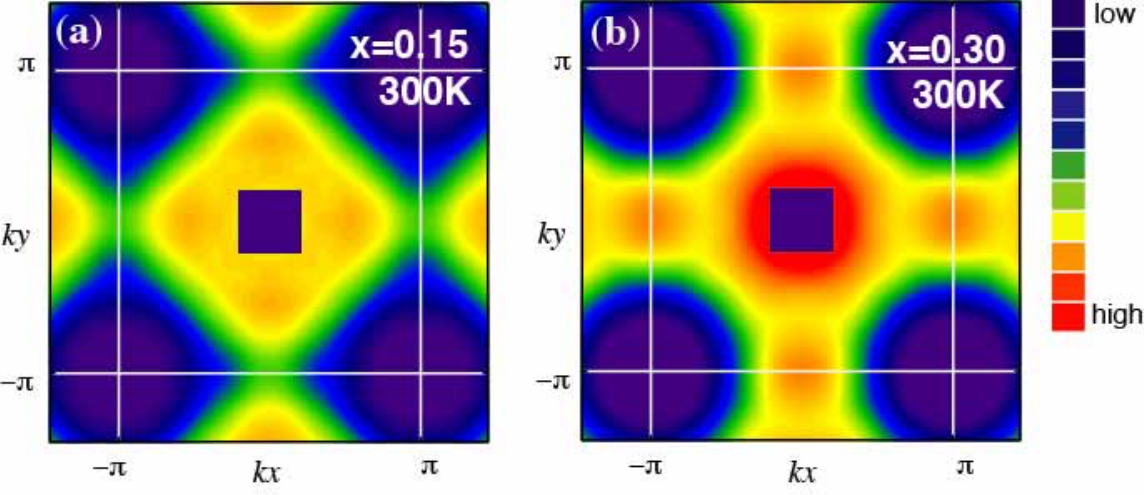}
\caption{(Color online) Images of the momentum distribution function $n(\vk)$ 
by high-resolution Compton scattering. 
Maps of $n(\vk)$ in the first Brillouin zone at 300~K for LSCO with $x=0.15$ (a) and 0.30 (b).  
The results for $x=0.08$ shown in \fig{nk}. 
The color scale represents the relative intensity.  
}  
\label{nk-doping300}
\end{figure}
%%%%%%%%%%%%%%%%%%%%%%%%%%%%%%%%%%%%%%%%%%%%%%%%%

Figures~\ref{nk-doping300}~(a) and (b) show maps of $n(\vk)$ at $x=0.15$ and $0.30$ 
measured by our Compton scattering. 
The obtained map at $x=0.15$ is similar to that for $x=0.08$ at 300~K [\fig{nk}~(a)], 
implying that the underlying FSs are also similar to each other. 
Note that the FS at $x=0.15$ is indeed very similar to that at $x=0.08$ 
in the conventional picture, too \cite{yoshida06}. 
In contrast, the map at $x=0.30$ is 
different from those for $x=0.08$ and $0.15$  at 300~K, suggesting different shapes of FSs at $x=0.30$. 
In fact, the conventional picture \cite{yoshida06} tells that 
the electron-like FS is realized at $x=0.30$ and the hole-like FS is at $x=0.08$ and 0.15. 

Under the condition of $\alpha=\alpha'' =0$, $\tilde{t}^{\,''}/\tilde{t}^{\,'}=-1/2$, and $\tilde{t}_z=0$ 
[see Eqs.~(\ref{xia})-(\ref{c-axis2})],  
we first determine the actual band parameters $\tilde{t}^{\,'}$ and $\tilde{t}^{\,''}$ 
to reproduce the FS proposed by ARPES \cite{yoshida12,yoshida06}. We obtain 
\be
\tilde{t}^{\,'} = -0.12 \tilde{t}, \quad  \tilde{t}^{\,''} = 0.06 \tilde{t} 
\ee
for both $x=0.15$ and $0.30$. The band parameters are different from those at $x=0.07$ [see \eq{tp-value}]. 
This doping dependence is well known in the tight-binding fit to ARPES data in La-based cuprates \cite{yoshida06}. 
While we shall also consider various $c$-axis dispersions [Eqs.~(\ref{c-axis}) and (\ref{c-axis2})] to perform 
a precise analysis as much as possible, the essential feature of our Compton scattering data is captured 
already without considering the $c$-axis dispersion. 

\subsubsection{Analysis of $n(\vk)$ for $x=0.15$}
\begin{figure} [t]
\centering
\includegraphics[width=10cm]{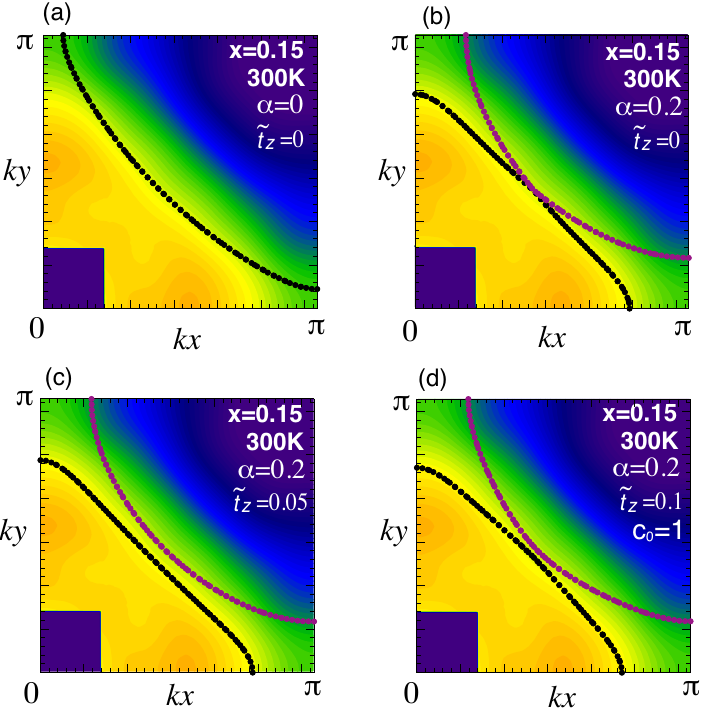}
\caption{(Color online) 
Interpretation of the observed maps of the momentum distribution function $n(\vk)$ 
for $x=0.15$ at 300~K. Expected FSs are superposed on the maps. 
(a) FS proposed by ARPES \cite{yoshida12}. 
(b) - (d) FSs with $\alpha=0.2$ in the nematic scenario. 
The $c$-axis dispersion is neglected in (b),  and Eqs.~(\ref{c-axis}) and (\ref{c-axis2}) 
for $k_z=0$ are considered in (c) and (d), respectively. 
}  
\label{nk0.15}
\end{figure}

In \fig{nk0.15}~(a), we superpose the FS proposed by ARPES \cite{yoshida12}. 
Along the FS, $n(\vk_{F})$ is almost constant 
in an extended region around $\vk=(0.43 \pi, 0.43 \pi)$. This nice agreement between ARPES and our Compton data is, 
however, broken around $\vk=(\pi,0)$ and $(0,\pi)$. This unsatisfactory aspect is resolved by considering 
FS deformation from the nematicity as shown in \fig{nk0.15}~(b). 
The inner FS is fully consistent with the map of $n(\vk)$. 
The outer FS is also almost consistent with our data, although a small region around $\vk=(0.43\pi, 0.43\pi)$ may not 
be so perfect. The agreement is improved when we introduce 
the $c$-axis dispersion in \eq{c-axis} as shown in \fig{nk0.15}~(c). 
The values of $\alpha$ and $\tilde{t}_z$ cannot be determined uniquely and 
a reasonably good agreement is achieved in $0.16 \lesssim \alpha \lesssim 0.25$ and 
$0 \lesssim \tilde{t}_{z} \lesssim 0.05$. 
Actually, a larger $\tilde{t}_z$ would be more consistent with our data, but the splitting of the hole- and 
electron-like FSs around $\vk=(0.43\pi,  0.43\pi)$ becomes sizable, which does not seem to be supported by 
ARPES data \cite{yoshida12}. 
A choice of a different $c$-axis dispersion [\eq{c-axis2}] yields essentially the same results 
[\fig{nk0.15}~(d)] even if we introduce $c_0=1$  
in \eq{c-axis2} as long as $0 \lesssim \tilde{t}_{z} \lesssim 0.1$. 
Compared to the case at $x=0.08$, a value of $\alpha$ becomes smaller at $x=0.15$. 
This smaller $\alpha$ comes from weaker nematic correlations 
and weaker lattice anisotropy with carrier doping.

 %%%%%%%%%%%%%%%%%%%%%% FIG. 18 %%%%%%%%%%%%%%%%%%%%%
%%%%%%%%%%%%%%%%%%%%%%%%%%%%%%%%%%%%%%%%%%%%%%%%%

\subsubsection{Analysis of $n(\vk)$ for $x=0.30$}
\begin{figure} [t]
\centering
\includegraphics[width=10cm]{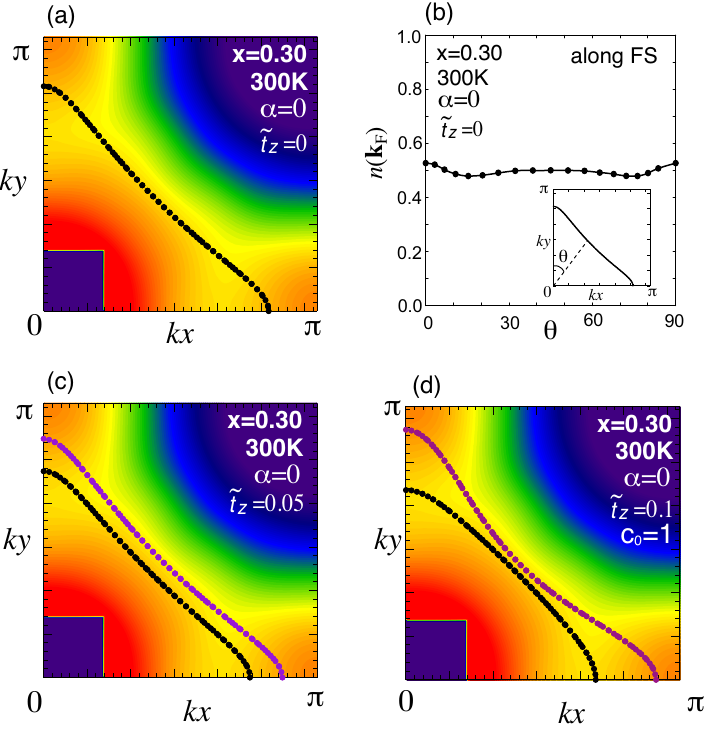}
\caption{(Color online) 
Interpretation of the observed maps of the momentum distribution function $n(\vk)$ 
for $x=0.30$ at 300~K. Expected FSs are superposed on the maps. 
(a) FS proposed by ARPES \cite{yoshida06}. 
(b) $n(\vk_F)$ along the expected FS shown in (a). The value of $n(\vk_F)$ 
is normalized to be $0.5$ at $\vk_F=(0.4\pi, 0.4\pi)$ 
and the angle $\theta$ is defined in the inset. 
(c) and (d) FSs in the presence of the $c$-axis dispersions Eqs.~(\ref{c-axis}) and (\ref{c-axis2}), respectively, 
for $k_z=0$.  
}  
\label{nk0.30}
\end{figure}

The FS proposed by ARPES \cite{yoshida06} is superposed on our map of $n(\vk)$ in \fig{nk0.30}~(a). 
Along the FS, $n(\vk_F)$ is almost constant, although the agreement seems less satisfactory around a region 
$\vk=(0.15\pi, 0.7\pi)$ and $(0.7\pi, 0.15\pi)$. 
To consider whether this is reasonably acceptable, we estimate the value of $n(\vk)$ 
along the FS shown in \fig{nk0.30}~(a) by assuming that $n(\vk_F)=0.5$ at $\vk_F=(0.4\pi, 0.4\pi)$. 
The result is shown in \fig{nk0.30}~(b). It turns out that $n(\vk_F)$ varies very sightly around $0.5$ along 
the expected FS, which indicates that the FS proposed by ARPES \cite{yoshida06} 
agrees with our Compton scattering data. 
This conclusion is reasonable because the spectral function $A(\vk, \omega)$ is rather sharp in ARPES 
at $x=0.30$  (Ref.~\onlinecite{yoshida06}) and 
thus there is not much room to invoke additional physics that ARPES potentially misses. 
In addition, our conclusion is consistent with the previous Compton scattering for $x=0.30$ (Ref.~\onlinecite{al-sawai12}). 
Inclusion of the $c$-axis dispersion such as Eqs.~(\ref{c-axis}) and (\ref{c-axis2}) does not alter our conclusion 
as seen in Figs.~\ref{nk0.30}~(c) and (d). Note that in general there should exist two FSs at a given $k_z$ 
in LSCO because the unit cell contains {\it two} CuO$_2$ planes reflecting 
the body-centered tetragonal crystal structure. 
If we invoke a larger $\tilde{t}_z$, a value of $n(\vk)$ tends to vary along the {\it outer} FS 
especially around $\vk=(\pi,0)$ and $(0,\pi)$. 
The reason why the momentum distribution function $n(\vk)$ 
is enhanced around $\vk=(\pi,0)$ and $(0,\pi)$ is left to further studies.

%%%%%%%%%%%%%%%%%%%%%% FIG. 16 %%%%%%%%%%%%%%%%%%%%%
%%%%%%%%%%%%%%%%%%%%%%%%%%%%%%%%%%%%%%%%%%%%%%%%%

While it seems unlikely to invoke the nematic physics at $x=0.30$, it is worth checking whether 
our data are indeed consistent with this expectation. 
We found that essentially the same FSs are obtained as those in Figs.~\ref{nk0.30}~(c) and (d) 
for the same band parameters except for a finite value of $\alpha \lesssim 0.05$. 
In this sense, our data cannot exclude possible nematicity also at $x=0.30$, but with a small $\alpha$ 
if there is. 

\end{document}